\documentclass{article}
\usepackage{my_style}
\usepackage[normalem]{ulem}
\pdfoutput=1

\begin{document}
\title{The threshold for quantum-classical correspondence is $D \sim \hbar^{\frac43}$}
\author[1]{Felipe Hern\'andez\footnote{felipeh@psu.edu}}
\author[2]{Daniel Ranard\footnote{dranard@caltech.edu}}
\author[3]{C.~Jess Riedel\footnote{jessriedel@gmail.com}}
\affil[1]{Department of Mathematics, Penn State University, State College, PA 16802, USA}
\affil[2]{Department of Physics, California Institute of Technology, Pasadena, CA 91125, USA}
\affil[3]{NTT Research, Inc., Physics \& Informatics Laboratories, Sunnyvale, CA 94085, USA}

\maketitle

\begin{abstract}
 In chaotic quantum systems, an initially localized quantum state can deviate strongly from the corresponding classical phase-space distribution after the Ehrenfest time $t_{\mathrm{E}} \sim \log(\hbar^{-1})$, even in the limit $\hbar \to 0$. Decoherence by the environment is often invoked to explain the persistence of the quantum-classical correspondence at longer timescales. Recent rigorous results for Lindblad dynamics with phase-space diffusion strength $D$ show that quantum and classical evolutions remain close for times that are exponentially longer than the Ehrenfest time whenever $D \gg \hbar^{\frac43}$, in units set by the classical Hamiltonian.  At the same time, some heuristic arguments have suggested the weaker condition $D \gg \hbar^{2}$ always suffices. Here we construct an explicit Lindbladian that demonstrates that the scaling $D \sim \hbar^{\frac43}$ is indeed the threshold for quantum-classical correspondence beyond the Ehrenfest time. Our example uses a smooth time-dependent Hamiltonian and linear Lindblad operators generating homogeneous isotropic diffusion. It exhibits an $\hbar$-independent quantum-classical discrepancy at the Ehrenfest time whenever $D \ll \hbar^{\frac43}$, even for $\hbar$-independent ``macroscopic'' smooth observables.  
\end{abstract}

\section{Introduction}

In chaotic quantum systems, the quantum dynamics can differ substantially from the  
corresponding classical dynamics after the Ehrenfest time $\teh\sim\log(S_H/\hbar)$, even as $\hbar/S_H \to 0$.  Here $S_H$ is a characteristic scale of the classical Hamiltonian $H$, with units of action. 
The discrepancy is due to coherent superpositions in the quantum dynamics, which have no classical analogue.
One conventional explanation for the emergence of classical mechanics in macroscopic 
systems invokes decoherence to destroy these superpositions by coupling with the external environment \cite{zeh1973quantum,kubler1973dynamics,zurek1981pointer, joos1985emergence,zurek1994decoherence,shiokawa1995decoherence, zurek1998decoherence,zurek1999why,perthame2004quantum} 
(cf.~\cite{ballentine1994inadequacy,fox1994chaos1,schlautmann1995measurement,casati1995comment,emerson2002quantum,wiebe2005quantum, kofler2007classical,zurek1995zurek,fox1994chaos2,brun1996quantum,kofler2008conditions}).  
However, the question remains of how to \textit{quantify} the strength of coupling necessary for classical dynamics to emerge.  That is, if decoherence is needed to ensure classicality as $\hbar/S_H \to 0$, \emph{how much do we need exactly}?

In the semiclassical limit, closed quantum systems may be modeled via a classical probability distribution on phase space, evolving under classical Hamiltonian flow. Markovian open systems are described by the Lindblad equation, which supplements the Schr\"odinger equation with Lindblad terms that arise due to the environment.  In the semiclassical limit, these Lindblad terms modify the corresponding classical dynamics, supplementing the classical Hamiltonian flow with diffusion on phase space~\cite{hernandez2023decoherence,hernandez2025classical, galkowski2025classical, li2025long}.  That is, the Lindblad equation corresponds to a classical Fokker-Planck equation. 
We characterize the strength of the Lindblad terms by the diffusion coefficient $D$, which also dictates the rate of decoherence. 
For instance, superposed wavepackets separated by a distance $\ell$ decohere after a time proportional to $\hbar^2/(D\ell^2)$ \cite{joos1985emergence}. Thus superpositions separated by distances larger than the ``decoherence length''
\begin{equation} \label{eq:elldec}
    \elldec \sim \hbar /\sqrt{D}
\end{equation}
decohere quickly relative to the Hamiltonian dynamics.

In Refs.~\cite{hernandez2023decoherence,hernandez2025classical}, it was
shown that the quantum-classical correspondence
persists far beyond the Ehrenfest time
so long as $D \gg D_H (\hbar/S_H)^{\frac43}$, where $D_H$ is a characteristic scale of the classical Hamiltonian $H$ with units of diffusion, independent of the actual diffusion $D$.
Henceforth we focus on parameters $\hbar$ and $D$, while omitting dimensional factors like $S_H$ and $D_H$, effectively working in units where they are set to one. (Cf.~Ref.~\cite{hernandez2023decoherence}, where the characteristic scales are handled explicitly.)  
By studying the regime $\hbar \ll 1$, we are studying Hamiltonians that lack microscopic structure: the function $H(x,p)$ varies only on scales much larger than $\hbar$. 
A paradigmatic example would be a heavy particle in a smooth potential, weakly scattered by lighter particles modeled as a bath \cite{joos1985emergence}.


More precisely, Refs.~\cite{hernandez2023decoherence,hernandez2025classical} consider quantum systems of possibly several continuous variables, initiated in a mixture of Gaussian states, and they compare (i) the trajectory of quantum states $\rho_t$ that solves the Lindblad equation, to (ii) the trajectory of classical probability densities $\fo_t(\alpha)$ that solves the corresponding Fokker-Planck equation, with phase space coordinate $\alpha=(x,p)$. They prove
\begin{equation}
	\label{eq:HRR-errbd}
	\left\lvert \Trace[\rho_t \hat{\gp}] - \int \fo_t(\alpha) \gp(\alpha)\diff \alpha \right\rvert < C t(\hbar^2 D^{-\frac32} +\hbar^{\frac12}) 
	(\|\hat{\gp}\|_{\mathrm{op}} + \|\gp\|_{L^\infty}),
\end{equation}
where $C$ is independent of $\hbar$.
The left-hand side compares the quantum expectation value $\Trace[\rho_t \hat{\gp}]$ of the quantum observable $\hat{\gp}$ with the corresponding classical expectation.  The latter is an integral over phase space of the ``classical observable'' $\gp(\alpha)$,
where $\hat{\gp}$ and $\gp(\alpha)$ are related by the Wigner-Weyl transform.\footnote{Refs.~\cite{hernandez2023decoherence,hernandez2025classical} moreover introduce an evolving mixture of Gaussian states, with both small trace-norm distance to $\rho_t$ and small $L^1$ distance to $\fo_t$. Ref.~\cite{li2025long} directly compares $\rho_t$ and the Weyl quantization $\hat{f}_t$ in trace-norm, for the regime $D \gg \hbar$. Ref.~\cite{galkowski2025classical} makes a comparison in Hilbert-Schmidt norm, for $D \gg \hbar^{\frac43}$.}

For $\hbar^2 D^{-\frac32} \ll 1$, or equivalently $D\gg\hbar^{\frac43}$, the right-hand side of Eq.~\eqref{eq:HRR-errbd} is small for long times.
More precisely, for $D > \hbar^p$ with $p < 4/3$, the error is small for times $t=O(\hbar^{-c})$ for some $c>0$, exponentially
longer than the Ehrenfest time\footnote{We refer to $\teh \sim \log \hbar^{-1}$ as the Ehrenfest time regardless of whether the system is chaotic. More precisely, $\teh \sim \lambda^{-1} \log \hbar^{-1}$ where $\lambda$ is the maximum \emph{instantaneous} Lyapunov exponent, or local expansion rate, controlled by second derivatives of $H$.}  $\teh \sim \log(\hbar^{-1})$.
In Refs.~\cite{hernandez2023decoherence,hernandez2025classical}, the question was raised as to whether the $D \sim \hbar^{\frac43}$ ``threshold'' was an artifact of the proof or indicated a real phenomenon.

On the one hand, numerical evidence \cite{toscano2005decoherence,wisniacki2009scaling} and heuristic arguments in Ref.~\cite{hernandez2023decoherence} indicate that the error between the quantum state and the closest classical approximation might indeed grow at rate $\hbar^2 D^{-\frac32}$, corresponding to a genuine threshold at $D\sim \hbar^{\frac43}$.  On the other hand, there are heuristic arguments that $D \gg \hbar^2$ is already sufficient for quantum-classical correspondence \cite{kolovsky1994remark, kolovsky1996condition, zurek2003decoherence, pattanayak2003parameter, gammal2007quantum}.  Crucially,  $D \gg \hbar^2$ yields a decoherence length $\elldec \ll 1$, suppressing macroscopic superpositions.  However, this does not automatically imply that the classical and quantum evolutions approximately match, and conflicting heuristics yield different conclusions.\footnote{Ref.~\cite{zurek2003decoherence}, Eq.~(5.55--5.66) and Ref.~\cite{kolovsky1994remark}, Eq.~(9), estimate the threshold for classical behavior by requiring that the leading quantum correction to $\partial_t W$ for the Wigner function $W$ is small pointwise \emph{relative} to the leading classical term. In our notation, this condition becomes  $D \gg \hbar^{2}$. In contrast, a heuristic presented in Ref.~\cite{hernandez2023decoherence} considers the \emph{absolute} size of the $\partial_t W$ correction, suggesting the stronger condition $D \gg \hbar^{\frac43}$ may be necessary. Separately, Ref.~\cite{kolovsky1996condition}, Eq.~(11), argues for $D \gg \hbar^{2}$ as sufficient for correspondence, but does so based on the condition that the \emph{macroscopically distinguishable} discrete paths in the Van Vleck–Gutzwiller open-system propagator decohere, i.e., that macroscopic superpositions are suppressed; as shown in the present paper, this does \emph{not} ensure actual correspondence between observables.} 

To complicate matters, one might expect a distinction when demanding a quantum-classical correspondence only for ``macroscopic'' observables, whereas the bound~\eqref{eq:HRR-errbd} treats more general observables.
By macroscopic observables, we mean smooth functions of phase space, such as $x$ or $p^2$, which do not depend on $\hbar$; these cannot probe microscopic oscillations in the wavefunction.  
In fact, even for \emph{closed} systems, it has been argued that for Hamiltonians without $\hbar$-scale structure, the quantum-classical correspondence may hold for macroscopic observables beyond the Ehrenfest time~\cite{gong2003chaos, emerson2001characteristics, wiebe2005quantum}, with error vanishing as $\hbar \to 0$.  
Pragmatically, the de Broglie wavelength associated with macroscopic superpositions is extremely small, e.g., $\hbar/(1\, \mathrm{kg\, m/s}) \sim  10^{-34}\,\mathrm{m}$, far smaller than a nucleus. 
Since interference on this scale seems infeasible to detect, one might suspect that an effective quantum-classical correspondence holds beyond the Ehrenfest time when restricted to $\hbar$-independent observables.
This view is supported by numerical results for specific systems~\cite{gong2003chaos, emerson2001characteristics, wiebe2005quantum}. 

A delicate interplay of time scales also casts doubt on whether the quantum-classical correspondence can break down for macroscopic observables, especially in the presence of noise.  Quantum-classical correspondence holds \emph{before} the Ehrenfest time, $t \ll \teh$,
by Egorov's theorem \cite{egorov1969canonical, zworski2022semiclassical, bambusi1999long,combescure1997semiclassical}.
Sensitivity to initial conditions appears necessary to see a breakdown around $t \sim \teh$, associated with chaos. On the other hand, for strongly chaotic systems, once an initially localized state becomes macroscopically delocalized at $t \sim \teh$, one expects it mixes within time $O(\teh)$ as well \cite{bonechi2000exponential}. 
(In phase spaces that are large or not fully chaotic, mixing may occur only later.)
After mixing, macroscopic observables take their equilibrium values, which exhibit quantum-classical correspondence as $\hbar \to 0$ \cite{simon1980classical,van2024gibbs}.  If the correspondence holds both before the Ehrenfest time and often somewhat after, one may wonder whether this allows any room for a quantum-classical breakdown---especially in the presence of noise that prevents macroscopic superpositions. 

We resolve these questions by demonstrating that, for a specific choice of smooth Hamiltonian $H$ and diffusion strength\footnote{Note that in other papers such as Ref.~\cite{hernandez2023decoherence}, $D$ is used to denote the diffusion matrix, while in this paper we always use a diffusion matrix proportional to the identity matrix, for our choice of $x$ and $p$ units, so that $D$ is just the non-negative scalar proportionality constant.}
$D$, the quantum-classical correspondence breaks down at the Ehrenfest time whenever $D\ll \hbar^{\frac43}$, and this breakdown manifests even for macroscopic observables.  The Hamiltonian we exhibit is quite simple, cubic in $x,p$ for a single variable.  
This ``counterexample'' is our main technical result.  It is a counterexample to the suggestion that $D \gg \hbar^2$ is sufficient for post-Ehrenfest correspondence; $D\gg \hbar^{\frac43}$ is sometimes necessary.

\subsection{Main result}

To state the main result, we establish some notation.  Given a time-dependent classical Hamiltonian function $H(t)$ on phase space,
we write $\HQ(t) = \Op_\hbar(H)(t)$
for its Weyl quantization and consider the open quantum system dynamics via the Lindblad equation with constant homogeneous isotropic diffusion,
\begin{equation}
\label{eq:Lindblad}
\partial_t \rho = -\frac{i}{\hbar} [\HQ(t),\rho] - \frac{D}{2\hbar^2} \left([\hat{x},[\hat{x},\rho]] +[\hat{p},[\hat{p},\rho]]\right).
\end{equation}
The corresponding Fokker-Planck equation is given by 
\begin{equation}
\label{eq:FP}
\partial_t \fo = \{\!\{H(t), \fo\}\!\}_{\rm PB} + \frac{D}{2} ( \partial_x^2 \fo + \partial_p^2 \fo),
\end{equation}
where $\{\!\{A,B\}\!\}_{\rm PB} = \partial_x A \partial_p B - \partial_p A \partial_x B$ is the Poisson bracket. 

\begin{theorem}
\label{thm:main-result}
There exists a smooth time-dependent Hamiltonian, e.g.,
\begin{align}
\HQ(t) = g_1(t)(\hat{x}\hat{p}+\hat{p}\hat{x}) +g_2(t) \hat{x}^3,
\end{align}
such that the following holds:
Let $\rho_t$ be the solution to the Lindblad equation~\eqref{eq:Lindblad}, with 
initial data $\rho_0 = \ket{0}\!\bra{0}$ given by the coherent state $\ket{0}$ centered at the origin in phase space. Let
$\fo_t$ solve the Fokker-Planck equation~\eqref{eq:FP} with $\fo_0$ being the Wigner function of $\rho_0$.  
For some smooth bounded observables, e.g., $\pf(\hat{p})=e^{-\hat{p}^2}$,
there exist $\hbar$-independent constants $C, c >0$ such that
\begin{align}\label{eq:main-result}
\left\lvert\Trace[\pf(\hat{p})\rho_T] - \int \pf(p) \fo_T(x,p) \diff x \diff p \right\rvert \geq c - C (1 + \log \hbar^{-1}) \hbar^{-\frac43}D
\end{align}
for some $T$ with $T = O(1+ \log \hbar^{-1})$.
\end{theorem}
\noindent In particular, when $D \leq \frac12 c C^{-1}(\log\hbar^{-1})^{-1}\hbar^{\frac43}$, the right-hand side of \eqref{eq:main-result} yields a strictly positive $\hbar$-independent lower bound.   
Thus, even for smooth bounded\footnote{As shown in the Appendix, one may also witness the discrepancy even in simple polynomials like $p^3$.} observables,  the threshold $D \sim \hbar^{\frac43}$ suggested by Eq.~\eqref{eq:HRR-errbd} for long-time quantum-classical correspondence cannot be relaxed (up to logarithmic factors): if $D \lsim \hbar^{\frac43}/\log(\hbar^{-1})$, there exist systems and smooth observables for which the quantum-classical discrepancy at Ehrenfest time is bounded below by an $\hbar$-independent constant.  In this sense, the results of Refs.~\cite{hernandez2023decoherence,hernandez2025classical} are asymptotically tight in their $D$-dependence, and $D\sim\hbar^{\frac43}$ is the correct boundary.

While we treat the system in Theorem~\ref{thm:main-result} as a counterexample to the possibility of improving the $D$-dependence of Eq.~\eqref{eq:HRR-errbd} of Refs.~\cite{hernandez2023decoherence,hernandez2025classical}, there are a few technical incongruities between the setting of Theorem~\ref{thm:main-result}  and the setting of the bounds in Refs.~\cite{hernandez2023decoherence,hernandez2025classical}.  We address these below:
\begin{enumerate}
\item Ref.~\cite{hernandez2025classical} assumes a bounded Hamiltonian.  However, because the example in Theorem~\ref{thm:main-result} is well-localized to a constant-size region in phase space throughout $t \in [0,T]$, it could be modified to use a bounded Hamiltonian (as we explain in the proof).
\item Ref.~\cite{hernandez2023decoherence}, unlike Ref.~\cite{hernandez2025classical}, specializes to the case that $H(x,p)=Y(p) + V(x)$ is a separable function of $x$ and $p$. The example in Theorem~\ref{thm:main-result} could be straightforwardly cast as a sequence of evolutions of this form, after performing a sequence of canonical transformations. 
\item The example in Theorem~\ref{thm:main-result} uses a time-dependent Hamiltonian.  However, it is straightforward to generalize  Eq.~\eqref{eq:HRR-errbd} and the other results of Refs.~\cite{hernandez2023decoherence,hernandez2025classical} to time-dependent Hamiltonians.\footnote{One may even show the time-dependent case merely by applying the time-independent case, after approximating the time-dependent evolution as a sequence of piecewise-constant evolutions.}   Alternatively, the time-dependent evolution of Theorem~\ref{thm:main-result} can be recast with a time-independent Hamiltonian, using an additional degree of freedom as a clock. 
\end{enumerate}

The Hamiltonian in Theorem~\ref{thm:main-result} is not chaotic in the sense usually
invoked in discussions of the Ehrenfest time, but it does feature an instability.  Indeed, the classical Hamiltonian $xp$ is equivalent to the inverted harmonic oscillator $p^2-x^2$ up to a canonical transformation.  This instability is what we use to achieve a breakdown of the quantum-classical correspondence at the relatively short time $\teh \sim \log(\hbar^{-1})$.  
Our dynamics, however, keeps the state effectively confined to a region of unit diameter in phase space, so long as the initial state is confined to a region of size $\hbar^{\frac12}$. In Section~\ref{sec:regimes}, we speculate whether the quantum-classical correspondence breaks down for more generic chaotic systems.

Finally, we remark that $D\sim\hbar^{\frac43}$ is only a ``sharp'' transition in the following asymptotic sense: if we scale $D$ as $D=\hbar^p$ and take $\hbar \to 0$, then $p=4/3$ marks a sharp change in behavior.  For any fixed nonzero $\hbar$ and fixed choice of Hamiltonian, varying $D$ does not lead to a sharp change in the dynamics.

\section{Construction}
\label{sec:motivation}
	
To understand the construction of the counterexample, we must first explain how the key factor $\hbar^2 D^{-\frac32}$  in~\eqref{eq:HRR-errbd} arises.  Let us sketch a toy version of the proof in Ref.~\cite{hernandez2023decoherence} for a system of a single variable.  In the sketch, we assume $D < \hbar$ to simplify some expressions.

Using the classical Hamiltonian $H$, first we fix a preferred ratio between position and momentum units in phase space.  
This determines a notion of phase-space length and a family of unsqueezed coherent states of width $\hbar^{\frac12}$ in both $x$ and $p$.   It allows us to treat the decoherence length $\elldec \sim \hbar D^{-\frac12}$ of Eq.~\eqref{eq:elldec} as a distance in phase space.
We assume the initial state can be expressed as a convex combination
\begin{align}
    \rho \approx \sum_j q_j \,\tauQ_{\alpha_j,\cov_j},
    \qquad
    q_j \ge 0,
    \qquad\ \sum_j q_j = 1, \qquad \cov_j^{\frac12} \lesssim \elldec,
    \label{eq:NTS-decomp}
\end{align}
where each $\tauQ_{\alpha_j,\cov_j}$ is a pure Gaussian state centered at $\alpha_j=(x_j,p_j)$ with covariance matrix $\cov_j>0$. The bounded covariance $ \cov_j^{\frac12} \lesssim \elldec$ means each Gaussian has width at most $\elldec$ in phase space. The approximation scheme proceeds as follows:

\begin{enumerate}
    \item \emph{Local quadratic evolution.} Given a mixture of Gaussian states $\tauQ_{\alpha_j,\cov_j}$ of limited squeezing as in \eqref{eq:NTS-decomp}, 
    approximate the dynamics of each $\tauQ_{\alpha_j,\cov_j}$ individually for a short time by taking a second-order Taylor approximation of the quantum Hamiltonian $\HQ$ centered at $\alpha_j$, which we call the Harmonic approximation $\HQ^{(\alpha_j)}$.  The reason to make this second-order approximation is that the quantum mechanical and classical evolution agree \textit{exactly} for such quadratic Hamiltonians. 
 Over the short evolution, we incur a \emph{harmonic error} (relative to the exact dynamics) due to the approximation, but we obtain a major benefit: the state remains \emph{exactly} Gaussian, albeit now generally  squeezed and (for $D\neq 0$) mixed.
    \item \emph{Decomposition into pure Gaussian states.} Although each Gaussian state is now squeezed and mixed, the diffusion ensures that the state never gets narrower than $\sim D^{\frac12}$ in any phase-space direction.
    Thus it may be expressed as a Gaussian mixture of \emph{less squeezed} pure Gaussian states, each of dimension $\sim \hbar D^{-\frac12} \times  D^{\frac12}$ (or even less squeezed), or width at most $\elldec$.
    Thus we obtain a mixture of Gaussian states again satisfying the assumptions of step (1).
    \item Repeat steps (1) and (2) until the final time.
\end{enumerate}
At the end, we obtain an approximation of the form \eqref{eq:NTS-decomp} to the exactly evolved quantum state.  
Furthermore, using the same mixture $q_j$ but now with corresponding classical Gaussians $\tauC_{\alpha_j,\cov_j}$ defines
a classical state which (for similar reasons) approximates the solution to the exact classical dynamics. The harmonic error rate (i.e., harmonic error per unit time) incurred by this approximation scheme is
\begin{align}
    \hbar^{-1}\left\|\left[\HQ-\HQ^{(\alpha_j)}, \tauQ_{\alpha_j,\cov_j}\right]\right\|_{\rm Tr}  \lsim \hbar^{-1} \elldec^3 \sim \hbar^2 D^{-\frac32}.
\end{align}
This explains the appearance of the factor $\hbar^2 D^{-\frac32}$.  The cubic exponent on $\elldec$ arises through Taylor's theorem because quantum and classical dynamics match exactly for quadratic Hamiltonians. 
This in turn accounts for the initially obscure exponent $4/3$ in the threshold $D\sim\hbar^{\frac43}$.

Working backwards, this argument also explains what must happen in order for a quantum-classical discrepancy to appear starting from  an initial coherent 
state $\rho_0$. To make our construction, we split up the quantum dynamics into three steps as
\begin{align}
\mcal{Q} := e^{\tau_3 \LLQ_3} e^{\tau_2 \LLQ_2} e^{\tau_1\LLQ_1}
\end{align}
and apply it to an isotropic coherent state $\rho_0=\ket{0}\bra{0}$ which is a Gaussian state $\tauQ_{0,\cov_0}$ 
with mean $0$ and covariance $\cov_0 \sim \diag(\hbar,\hbar)$.  The $\LLQ_i$ are built with Hamiltonians $\HQ_i$ and universal diffusion $D$. The classical counterpart of the evolution is
\begin{align}
\mcal{C} := e^{\tau_3 \LLC_3} e^{\tau_2 \LLC_2} e^{\tau_1\LLC_1}
\end{align}
applied to the classical Gaussian probability density $\fo_0 = \tauC_{0,\cov_0}$.  
\begin{figure}[t]
\centering
	\includegraphics[width=0.50\linewidth]{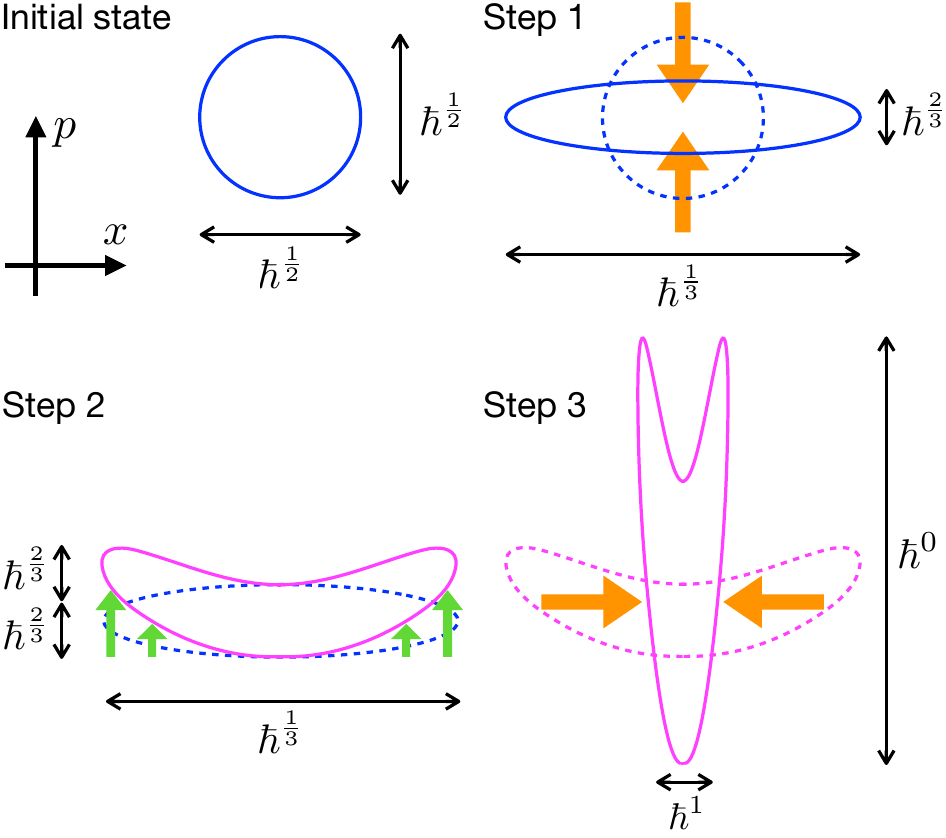}
	\caption{
        The dynamics of our counterexample in phase space. Quantum and classical Gaussian states are uniquely represented by their covariance ellipse. We assume an initial pure unsqueezed Gaussian state (blue circle) with 
        dimensions $\hbar^{\frac12}\times \hbar^{\frac12}$, i.e., covariance matrix $\cov_0 \sim \diag(\hbar,\hbar)$.
        In Step 1 we squeeze with $\HC_1$ for time $\tau_1$ to produce a Gaussian state (blue ellipse) with 
        dimensions $\hbar^{\frac13}\times\hbar^{\frac23}$.  
        In Step 2 we apply the nonlinearity $\HC_2$ for unit time, producing a non-Gaussian state (pink banana) whose variance has changed only by order unity. The quantum (but not classical) momentum distribution now contains oscillations induced by interference between opposite ends of the banana, but confined to the mesoscopic scale. In Step 3 we squeeze in the opposite direction to produce the final state (squashed pink banana), bringing the $p$ distribution to the macroscopic scale (see Fig.~\ref{fig:p-distribution}) with 
        dimensions of order $\hbar^1\times \hbar^0$.
	}
	\label{fig:ellipse-to-banana}
\end{figure}
The evolution is designed to exactly 
reverse-engineer a quantum-classical discrepancy as follows:
\begin{enumerate}
    \item The first step of the dynamics produces a state $\rho_1 = e^{\tau_1\LLQ_1}\rho_0$ that has coherence over a length scale $\ell_* \sim \hbar^{\frac13}$ (which is $\elldec \sim \hbar/D^{\frac12}$ when at the threshold $D\sim\hbar^{\frac43}$).   This is done by applying the Hamiltonian $\HQ_1 :=\frac12 (\hat{x}\hat{p}+\hat{p}\hat{x})$
    for time $\tau_1 = \frac16 \log \frac2\hbar$, which approximately constructs a pure Gaussian state, $\rho_1 \approx \tauQ_{0,\cov_1}$, with $\cov_1 = \diag(\hbar^{\frac23},\hbar^{\frac43})/2$ (with corresponding classical state $\fo_1 = e^{\tau_1 \LLC_1}\fo_0 \approx \tauC_{0,\cov_1}$).  That is, $\cov_1$ is stretched to scale $\hbar^{\frac13}$ along the $x$ coordinate.  
    \item In the second step $\rho_2=e^{\tau_2\LLQ_2}\rho_1$ (resp.\ $\fo_2=e^{\tau_2\LLC_2}\fo_1$), we apply the Hamiltonian $\HQ_2 := -\frac13\hat{x}^3$  for unit time $\tau_2 = 1$, so that the error term appearing in the Moyal bracket is nonvanishing.\footnote{Here any smooth Hamiltonian with nonvanishing third derivative clearly suffices, but taking $-\frac13\hat{x}^3$ simplifies the computation somewhat.} 
    \item The quantum-classical discrepancy between $\rho_2$ and $\fo_2$ is only apparent at momentum scales $\hbar^{\frac23}$.  This is magnified in the third step by applying Hamiltonian $\HQ_3 := -\frac12(\hat{x}\hat{p}+\hat{p}\hat{x}) = -\HQ_1$ for time $\tau_3 = \frac23 \log \frac2\hbar$ to obtain $\rho_3=e^{\tau_3\LLQ_3}\rho_2$ (resp.\ $\fo_3=e^{\tau_3\LLC_3}\fo_2$) which have a discrepancy apparent at macroscopic scales.
\end{enumerate}
These three steps are illustrated in Figure~\ref{fig:ellipse-to-banana}.  As a result of the construction, we obtain states 
which satisfy (assuming $D \ll \hbar^{\frac43}$)
\begin{align}\label{eq:qc-discrepancy}
\left \lvert \Trace[\mcal{Q}[\rho_0] \hat{\gp}] - \int (\mcal{C}\fo_0)(\alpha) \gp(\alpha)\diff\alpha \right \rvert > c 
\end{align}
for some $p$-diagonal observable
$\gp(x,p) = \pf(p)$ and $\hat{\gp} = \Op[\gp]= \pf(\hat{p})$ and a corresponding $\hbar$-independent constant $c$. One simple choice is $\pf(p) = \exp(-p^2)$, which is shown in the Appendix to allow $c \approx 0.06412$. 

\begin{figure}[t]
    \centering
	\includegraphics[width=0.5\linewidth]{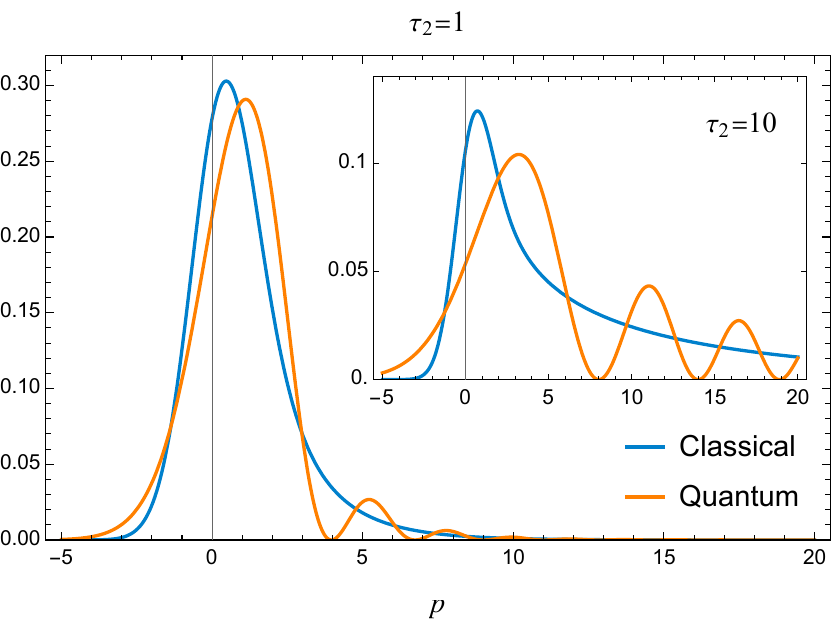}
	\caption{
		The classical (blue) momentum distribution 
        $\cmo_3(p) = \int\! \fo_3(x,p)\, \diff x$ 
        vs.\ the quantum (orange) momentum distribution 
        $\qmo_3(p) = |\hat{\psi}_3 (p)|^2$ 
        at the end of Step 3 of our protocol, plotted for the closed system, which approximates the open system when decoherence is weak, $D \ll \hbar^{\frac43}$. 
        We choose 
        $\tau_1=\tfrac{1}{6}\log (2/\hbar)$ and $\tau_3=\tfrac{2}{3}\log (2/\hbar)$,
        ensuring the distributions are independent of $\hbar$, and compare $\tau_2 = 1$ (main) with $\tau_2 = 10$ (inset).  The interference-induced oscillations in the quantum (but not classical) momentum distribution for positive $p$ are produced by interference between the two ends of the squashed banana in Fig.~\ref{fig:ellipse-to-banana}.  These oscillations already exist in the momentum distribution at the end of Step 2, but at a mesoscopic scale. Evidence of the interference is magnified to the macroscopic scale in Step 3 at the cost of allowing it to decohere.
	}
	\label{fig:p-distribution}
\end{figure}

The proof proceeds by showing that the quantum and classical observables can both be well approximated by $\hbar$-independent explicit integrals involving special functions. The classical $p$-distribution at the final time $T=\tau_1+\tau_2+\tau_3$ is given by a parabolic cylinder function, while the quantum $p$-distribution by a squared Airy function.  These are plotted in Fig.~\ref{fig:p-distribution}.  Since both distributions are $\hbar$-independent, they can be distinguished by smooth observables. 

In the remainder of the main body of the paper, we explain how this approximation is made
in more detail.  The squeezing and stretching steps involving $\LLQ_1$ and $\LLQ_3$ are discussed in Section~\ref{sec:squeezing}.  The effect of $\LLQ_2$ and the appearance of the special functions are presented in Section~\ref{sec:cubic} to obtain the result \eqref{eq:qc-discrepancy}.  A complete and more detailed calculation is deferred to the Appendix.

\section{Squeezing the observable and state}
\label{sec:squeezing}
In this section we deal with the squeezing and stretching steps $\LLQ_1$ and $\LLQ_3$, which are generated by Hamiltonians
$\HQ_1 = \frac12(\hat{x}\hat{p}+\hat{p}\hat{x})$ and $\HQ_3=-\HQ_1$, respectively.  It is convenient for us to work
with the stretching step by duality (the Heisenberg picture), so that it acts on the observable rather than the state, as follows:
\[
 \Trace[\mcal{Q}[\rho_0] \pf(\hat{p})] = \Trace[e^{\tau_3 \LLQ_3}e^{\tau_2 \LLQ_2}e^{\tau_1 \LLQ_1}[\rho_0] \pf(\hat{p})]
= \Trace[e^{\tau_2 \LLQ_2}e^{\tau_1 \LLQ_1}[\rho_0] e^{\tau_3\LLQ^*_3}\pf(\hat{p})].
\]
We have the following bounds on the error between the open and closed quantum dynamics:
\begin{align}
\label{eq:quantum-squeeze}
\|e^{\tau_1\LLQ_1}[\tauQ_{0,\cov_0}] - \tauQ_{0,\cov_1}\|_{\rm Tr} &\lsim D\hbar^{-\frac43} \tau_1 \\
\label{eq:quantum-stretch}
\|e^{\tau_3\LLQ_3^*}[\pf(\hat{p})] - \pf(e^{\tau_3}\hat{p})\|_{\rm op} &\lsim D\hbar^{-\frac43} \tau_3.
\end{align} 
Hence $\rho_1 = e^{\tau_1\LLQ_1}[\tauQ_{0,\cov_0}]$ is close to the pure Gaussian state $\tauQ_{0,\cov_1}$ resulting from the closed-system evolution, and likewise the Heisenberg-evolved observable $e^{\tau_3\LLQ_3^*}[\pf(\hat{p})]$ is close to the unitarily squeezed observable $\pf(e^{\tau_3}\hat{p})$.
Analogously, the
classical dynamics satisfy the estimates
\begin{align}
\label{eq:classical-squeeze}
\|e^{\tau_1\LLC_1}[\tauC_{0,\cov_0}] - \tauC_{0,\cov_1}\|_{L^1} &\lsim D\hbar^{-\frac43} \tau_1\\
\label{eq:classical-stretch}
\|e^{\tau_3\LLC_3^*}[\pf(p)] - \pf(e^{\tau_3}p)\|_{L^\infty} &\lsim D\hbar^{-\frac43}\tau_3.
\end{align} 
We choose $\tau_1 = \frac16 \log\frac2\hbar$ and $\tau_3=\frac23\log\frac2\hbar$, so that the errors above have magnitude
on the order $D\hbar^{-\frac43}$ (up to factors $\log(\hbar^{-1})$ we omit here), matching the right 
hand side in~\eqref{eq:main-result}.

\subsection{Squeezing states}
\label{sec:state-squeezing}
The closed
system dynamics up to time $s$ send $\tauQ_{0,\cov_0}$ to $\tauQ_{0,\cov(s)}$ with 
$\cov(s) = \diag(e^{2s}\hbar, e^{-2s}\hbar)/2$.  We therefore take  $\tau_1 = \frac16 \log\frac2\hbar$ so 
that $\cov_1 \sim \diag(\hbar^{\frac23}, \hbar^{\frac43})$.  We compare the dynamics of the
open-system evolution with diffusion $D$ to that of the closed-system dynamics using the Duhamel principle,
which in this case reads
\begin{align*}
e^{\tau_1\LLQ_1}[\tauQ_{0,\cov_0}] 
&= \tauQ_{0,\cov_1} - \frac{D}{2\hbar^2} \int_0^{\tau_1}  e^{(\tau_1-s)\LLQ_1}
\left([\hat{x},[\hat{x},\tauQ_{0,\cov(s)}]] +[\hat{p},[\hat{p},\tauQ_{0,\cov(s)}]]\right) \diff s.
\end{align*}
Using the triangle inequality in the trace norm, we therefore estimate
\begin{align*}
\|e^{\tau_1\LLQ_1}[\tauQ_{0,\cov_0}] -\tauQ_{0,\cov_1}\|_{\rm Tr} 
\lsim \frac{D}{\hbar^2} \int_0^{\tau_1}  \left\|[\hat{x},[\hat{x},\tauQ_{0,\cov(s)}]]\right\|_{\rm Tr} +
\left\|[\hat{p},[\hat{p},\tauQ_{0,\cov(s)}]]\right\|_{\rm Tr} \diff s.
\end{align*}
The larger term on the right is the double-commutator against $\hat{x}$, which has scale 
$e^{2s}\hbar \lesssim \hbar^{\frac23}$ for $s\le\tau_1$, and this completes the explanation of~\eqref{eq:quantum-squeeze}.

On the classical side the calculation is completely analogous, and we obtain
\begin{align*}
\|e^{\tau_1\LLC_1}[\tauC_{0,\cov_0}] -\tauC_{0,\cov_1}\|_{L^1} 
\lsim D \int_0^{\tau_1}  (\|\partial_x^2 \tauC_{0,\cov(s)}\|_{L^1}+\|\partial_p^2\tauC_{0,\cov(s)}\|_{L^1}) \diff s.
\end{align*}
The larger term above is the second derivative $\partial_p^2$, which contributes
$e^{2s}\hbar^{-1} \lesssim \hbar^{-\frac43}$.  Thus we obtain~\eqref{eq:classical-squeeze}.

\subsection{Squeezing observables}
Under the closed system evolution $\hat{\mcal{H}}_3$, 
we have $e^{\tau_3 \hat{\mcal{H}}_3^*} [\pf(\hat{p})] = \pf(e^{\tau_3}\hat{p})$. Thus applying the Duhamel formula as above, 
\begin{align*}
e^{\tau_3 \LLQ_3^*}[\pf(\hat{p})]
&= \pf(e^{\tau_3}\hat{p}) - \frac{D}{2\hbar^2} \int_0^{\tau_3} e^{(\tau_3-s) \LLQ_3^*} [\hat{x},[\hat{x},\pf(e^s \hat{p})]] \diff s \\
&= \pf(e^{\tau_3}\hat{p}) + \frac{D}{2} \int_0^{\tau_3} e^{(\tau_3-s) \LLQ_3^*}[e^{2s} \pf''(e^s \hat{p})] \diff s.
\end{align*}
By the triangle inequality, it therefore follows that 
\[
\|e^{\tau_3 \LLQ_3^*}[\pf(\hat{p})] - \pf(e^{\tau_3}\hat{p})\|_{\rm op}
\leq  D \|\pf''\|_{L^\infty}  \int_0^{\tau_3} e^{2s} \diff s  \leq   D e^{2\tau_3} \|\pf''\|_{L^\infty}.
\]
In particular, taking $\tau_3 = \frac23 \log \frac{2}{\hbar}$, we obtain
\begin{align}
\|e^{\tau_3 \LLQ_3^*}[\pf(\hat{p})] - \pf(e^{\tau_3} \hat{p})\|_{\rm op} = O(D \hbar^{-\frac43}),
\end{align}
which is~\eqref{eq:quantum-stretch}.
Exactly the same calculation on the classical side establishes~\eqref{eq:classical-stretch}.

\section{The Effect of the Cubic Hamiltonian}
\label{sec:cubic}
We are now in a position to compute the effect of the cubic potential,\footnote{One can use here instead any smooth and bounded $V(\hat{x})$ which vanishes to second order at $0$ and has a non-vanishing third derivative, using a Taylor approximation to estimate the error against the choice $\hat{x}^3$.}
$\HQ_2 = -\frac13\hat{x}^3$.
Combining the results of the previous section,
we have
\begin{equation}
\label{eq:quantum-integral}
\begin{split}
\Trace[\mcal{Q}[\rho_0] \pf(\hat{p})] 
&= \Trace[e^{\tau_2 \LLQ_2}e^{\tau_1 \LLQ_1}[\rho_0] e^{\tau_3\LLQ^*_3}[\pf(\hat{p})]] \\
&= \Trace[e^{\tau_2 \LLQ_2}[\tauQ_{0,\cov_1}] \pf(\hbar^{-\frac23} \hat{p})] + O(D\hbar^{-\frac43})
\end{split}
\end{equation}
and similarly
\begin{equation} 
\label{eq:classical-integral}
\int \mcal{C}[\fo_0](x,p)\pf(p) \diff x\diff p
= \int e^{\tau_2\LLC_2}[\tauC_{0,\cov_1}](x,p) \pf(\hbar^{-\frac23}p)\diff x \diff p + O(D\hbar^{-\frac43}).
\end{equation}
The quantum expectation value can be computed as an $\hbar$-independent integral involving an Airy
function, and the classical expectation value involves a parabolic cylinder function.  See Figure~\ref{fig:p-distribution} for an illustration.  In the remainder of
this section we see how these special functions arise from~\eqref{eq:quantum-integral} and~\eqref{eq:classical-integral}.

First, we work with~\eqref{eq:quantum-integral}.
Using exactly the same calculation as in Section~\ref{sec:state-squeezing}, we have 
\[
\|e^{\tau_2 \LLQ_2}[\tauQ_{0,\cov_1}] - e^{i\tau_2 \hat{x}^3/3\hbar} \tauQ_{0,\cov_1} e^{-i\tau_2 \hat{x}^3/3\hbar}\|_{\rm Tr} 
=O( D \hbar^{-\frac43})
\]
for any $\hbar$-independent $\tau_2$.
The second term inside the norm can be expressed explicitly:  it is the pure state with wavefunction
\[
\psi_2(x) := \pi^{-\frac14} \hbar^{-\frac16} e^{i\tau_2 x^3 /3\hbar} e^{-x^2/(2\hbar^{\frac23})}.
\]
In this case, $\psi_2(x) = \hbar^{-\frac16} \psi_3(x/\hbar^{\frac13})$ where   
$\psi_3(y) = \pi^{-\frac14} e^{i\tau_2 y^3/3} e^{-y^2/2}$ is the pure state that would be obtained after the third step with closed-system ($D=0$) dynamics.
Therefore, on the quantum side we have
\begin{align*}
\Trace[\mcal{Q}[\rho_0] \pf(\hat{p})] 
&= \int |\mcal{F}_\hbar[\psi_2](p)|^2 \pf(\hbar^{-\frac23} p) 
\diff p\\
&= \hbar^{-\frac23} \int |\hat{\psi}_3(\hbar^{-\frac23}p)|^2
\pf(\hbar^{-\frac23}p) \diff p \\
&= \int |\hat{\psi}_3(p)|^2 \pf(p)\diff p.
\end{align*}
Above, $\mcal{F}_\hbar$ is the semiclassical Fourier
transform, scaled by $\hbar$.  
The Fourier transform $\hat{\psi}_3(p)$ is given by an 
Airy function, and indeed for the $\hbar$-independent but otherwise arbitrary choice $\tau_2 =1$ we can infer that the final momentum distribution for $D \ll \hbar^{\frac43}$  is  
\begin{align}\label{eq:body-quantum-momentum-distribution}
	\qmo(p)  := |\hat{\psi}_3 (p)|^2 =  2^{\frac{1}{6}} \pi^{\frac12}  
	\exp \left[\frac{1-6 p }{12 }\right]
	\mathrm{Ai}\left(\frac{1-4 p}{2^{\frac83} }\right)^2.
\end{align}

On the other hand, on the classical side~\eqref{eq:classical-integral} we have that the closed-system 
evolution by $\LLC_2$ is transport by the map $(x,p) \mapsto (x,p+\tau_2 x^2)$. 
Using the Duhamel formula again, we obtain 
\begin{align}
e^{\tau_2 \LLC_2}[\tauC_{0,\cov_1}](x,p)
= \tauC_{0,\cov_1}(x,p-\tau_2 x^2) + \frac12 D
\int_0^{\tau_2} e^{(\tau_2-s)\LLC_2} [
(\partial_{xx} + \partial_{pp}) \tauC_{0,\cov_1}(x,p-s x^2)] \diff s.
\end{align}
The second derivatives are of order $\hbar^{-\frac43}$ because $\tauC_{0,\cov_1}$ is squeezed to width $\hbar^{\frac23}$ in the $p$ direction.
Therefore,
\begin{align*} 
\|e^{\tau_2 \LLC_2}[\tauC_{0,\cov_1}](x,p) - \tauC_{0,\cov_1}(x,p-\tau_2 x^2)\|_{L^1} = O(D\hbar^{-\frac43}),
\end{align*} 
so the classical observable can be estimated 
as follows:
\begin{align*} 
\int e^{\tau_3 \LLC_3} e^{\tau_2 \LLC_2} e^{\tau_1 \LLC_1}[\fo_0](x,p)
\pf(p) \diff x\diff p 
&= 
\int \tauC_{0,\cov_1}(x,p-\tau_2 x^2) \pf(\hbar^{-\frac23} p) \diff x\diff p
+ O(D\hbar^{-\frac43}).
\end{align*}
Expanding out the definition of $\tauC_{0,\cov_1}$, we obtain
\begin{align*}
\int (\mcal{C}\fo_0)(x,p) \pf(p)\diff x \diff p &=
\int \tauC_{0,\cov_1}(x,p-\tau_2x^2) \pf(\hbar^{-\frac23} p) \diff x\diff p \\
&= \int \left(\int \exp(-x^2/\hbar^{\frac23} - (p-\tau_2x^2)^2/\hbar^{\frac43}) \diff x\right)
\pf(\hbar^{-\frac23}p) \diff p \\
&=: \int \cmo(p) \pf(p)\diff p
\end{align*}
where, after a rescaling $x' = \hbar^{-\frac13}x$ and $p'=\hbar^{-\frac23}p$, we infer the final classical momentum distribution
\begin{align}\label{eq:body-classical-momentum-distribution}
	\cmo(p) &= 
	\frac{1}{2 \sqrt{\pi}}\,
	\exp\!\left[-\frac{p^2}{2}+\frac{1}{4}\left(p -\frac12 \right)^2\right]\,
	\pcf_{-\frac12}\!\left(\frac12  - p\right) ,
\end{align}
Here, $\pcf_{\ell}(z) = \Gamma(-\ell)^{-1}e^{-z^{2}/4}
\int_{0}^{\infty} e^{-z s - s^{2}/2} \, s^{-\ell-1} \, ds$ is a parabolic cylinder function for $\mathrm{Re}\, \ell <0$.
This differs from the corresponding quantum distribution 
function \eqref{eq:body-quantum-momentum-distribution}, which is also $\hbar$-independent. As illustrated in Fig.~\ref{fig:p-distribution}, these two distributions can be easily distinguished with smooth observables, e.g., $\pf(p) = \exp(-p^2)$, yielding Theorem~\ref{thm:main-result}.

\section{Discussion}\label{sec:regimes}
If $D \sim \hbar^{2}$ is not the boundary of the quantum-classical correspondence, what is it?  It is useful to think in terms of the decoherence length $\elldec \sim \hbar / \sqrt{D}$, which is expected to characterize the maximum length scale of persistent superpositions.  Then  $D \sim \hbar^{2}$ means $\elldec \sim 1$ and  we have at least three regimes:
\begin{itemize}
    \item \textbf{Macroscopic coherence} ($D \ll \hbar^{2}$): Decoherence is negligible, $\elldec \gg 1$, and the system is essentially closed in the classical limit $\hbar \to 0$. Beyond the Ehrenfest time, quantum superpositions generically form on macroscopic scales.
    \item \textbf{Mesoscopic coherence} ($\hbar^2 \ll D \ll \hbar^{\frac43}$): While macroscopic superpositions are suppressed, distinctly quantum behavior (superposition and possibly quantum computation) takes place below the decoherence length $\elldec \sim \hbar/\sqrt{D}$ in phase space.  This mesoscopic scale  is simultaneously much smaller than the macroscopic scale $\hbar^0$ and much larger than the microscopic scale $\hbar^{\frac12}$. As shown in our example, interference effects at the mesoscopic scale can be amplified to produce a macroscopic imprint, even when using a Hamiltonian that varies only at the macroscopic scale.
 
    \item \textbf{Quantum-classical correspondence} ($\hbar^{\frac43}  \ll D $): When the initial state is a 
    pure Gaussian state, or a mixture of such states, the evolution remains well-approximated by a mixture of such states, whose centroids flow along classical trajectories and diffuse about them with strength $D$~\cite{hernandez2023decoherence,hernandez2025classical}.  If $D = o(\hbar^0)$, then diffusion vanishes in the classical limit $\hbar\to 0$, so noiseless classical mechanics is recovered. 
    
     When the initial state is arbitrary, we expect the situation is more complicated. If $D \ll \hbar$, then $\elldec \gg \hbar^{\frac12}$, and an initial superposition of coherent states separated by distance $\sim \hbar^{\frac12}$ in phase space will only decohere slowly.  Therefore, for the case of $D \ll \hbar$ and \emph{arbitrary} initial states, we only conjecture a correspondence for smooth observables. In contrast, for $D \gg \hbar$, we speculate that arbitrary initial states may decohere sufficiently after an $o(1)$ time to yield a subsequent correspondence for all observables.
\end{itemize}

We wonder how generically the quantum-classical breakdown demonstrated by Theorem~\ref{thm:main-result} appears for other systems with $D \lesssim \hbar^{\frac43}$.  On one hand, the particular example in Theorem~\ref{thm:main-result} uses a fine-tuned initial state, sitting precisely at an unstable point; if it were displaced by a distance $\gg \hbar^{\frac12}$ in phase space, the state would not remain confined within a unit box (it would ``escape the laboratory'').
On the other hand, we speculate that for chaotic systems, a breakdown may occur regardless of where the initial coherent state is placed.  Generically it would stretch exponentially (though typically confined in phase space by energetic constraints), eventually squeezed to dimension $\hbar^{\frac23} \times \hbar^{\frac13}$, completing the analog of our Step 1 (Figure~\ref{fig:ellipse-to-banana}).  Then nonlinearities would produce small quantum-classical discrepancies, the analog of our Step 2.  Finally, these could be amplified to the macroscopic scale, the analog of our Step 3.  

Step 3 of our construction requires stretching the state in the opposite direction of Step 1: our time-dependent Hamiltonian involves an inversion of the stable and unstable directions of the flow.  If restricted to a time-independent Hamiltonian flow which is Anosov, and hence with invariant stable and unstable directions at every point, the analog of Step 3 is prohibited. We conjecture the threshold for quantum-classical correspondence is different for Anosov flows, perhaps at $D\gg \hbar^2$.  

One could also wonder whether the behavior exemplified in  Theorem~\ref{thm:main-result} is robust to noise or uncertainty in the initial condition, corresponding to a mixed initial state.  Likewise, one might consider noise in the Hamiltonian. We leave such questions to future work.

\section{Acknowledgments}
FH is supported by the NSF under award DMS-2303094.
 DR acknowledges support by the Simons Foundation under grant 376205. 

\newpage

\appendix

{\centering \section*{Appendix}}

In Appendix~\ref{sec:set-up}, we summarize the counterexample and set up the basic ingredients. In Appendix~\ref{sec:pure-evolution}, we solve the evolution for the case of a closed system (i.e., $D=0$).  In Appendix~\ref{sec:diffusion-error}, we bound the errors introduced by ignoring diffusion. In Appendix~\ref{sec:overall-bound}, we conclude with our overall bound.

\section{Set up}\label{sec:set-up}

We consider the quantum Lindblad evolution 
\begin{align}\label{eq:lindblad-app}
	\frac{\partial\rhoo}{\partial t} = \LLQ[\rhoo] &= -\frac{i}{\hbar}[\HQ,\rhoo]+\frac{1}{\hbar}\sum_i \left[ \LQ{i} \rhoo \LQ{i}^\dagger - \frac{1}{2}\left\{\LQ{i}^\dagger \LQ{i},\rhoo\right\}\right]\\
    &= -\frac{i}{\hbar} [\HQ,\rhoo] - \frac{D}{2\hbar^2} \left([\hat{x},[\hat{x},\rho]] +[\hat{p},[\hat{p},\rho]]\right).
\end{align}
with the two Lindblad operators $\{\LQ{i}\}_i = \{\sqrt{D/\hbar} \, \hat x, \sqrt{D/\hbar} \, \hat p\}$ producing uniform diffusion. This corresponds to the classical Fokker-Planck evolution  
\begin{align}\label{eq:fokker-planck}
	\frac{\partial\fo}{\partial t} = (\partial_x \HC)(\partial_p \fo)-(\partial_p \HC)(\partial_x \fo) + \frac{1}{2} D (\partial^2_x \fo + \partial^2_p \fo).
\end{align}
We want to compare these evolutions initialized in states with the same phase space distribution and governed by the same (specifically chosen) time-dependent Hamiltonian $H(t)$ (using the Wigner function $\Wo$ and the Wigner-Weyl quantized Hamiltonian $\HQ(t)$, respectively, in the quantum case).  In particular, our initial state will be an $\hbar$-scale uncertainty-principle-saturating Gaussian for both the quantum and classical systems. As it will simplify our expressions, let $h :=\hbar/2$ so that the uncertainty principle is saturated when $\sigma_x \sigma_p = h$ and in particular the initial coherent state enjoys $\sigma_x = \sigma_p = h^{\frac12}$.

Assuming
\begin{align}
    D=o\!\left(\frac{h^{\frac43}}{1+\log h^{-1}}\right),
\end{align}
we will show that after an amount of time $T = O(1+\log h^{-1})$, the quantum momentum distribution $\qmo_t(p) := \langle p|\rhoo_t |p\rangle$ and the classical momentum distribution $\cmo_t(p) := \int \!\diff x \, \fo_t(x,p)$ are distinguishable even as $\hbar\to 0$.  Specifically,
\begin{align}
\|\qmo_T-\cmo_T\|_{L^1} \ge \bar{c} - C\frac{1+\log h^{-1}}{h^{\frac43}} D 
\end{align}
for $\hbar$-independent constants $\bar{c},C$.  Furthermore, these distributions differ at low frequencies, i.e., they can be distinguished by smooth bounded observables diagonal in $p$ like $\hat{\gp}_0 :=  \exp(-\hat{p}^2)$:
\begin{align}
|\langle \hat{\gp}_0\rangle_{\rhoo_T} - \langle \gp_0\rangle_{\fo_T}| \ge c_0-C \frac{1+\log h^{-1}}{h^{\frac43}} D
\end{align}
where again $c_0$ is independent of $\hbar$.
For concreteness, we obtain $\bar{c} \approx 0.2726, C \approx 6.550, c_0 \approx 0.06412$ for our specific choice of $\HC(t)$.

\subsection{Initial state}

We will compare the quantum and classical evolution starting from an initial probability distribution $\fo_0$ and an initial coherent state whose Wigner function $\Wo_0$ coincides with $\fo_0$:
\begin{align}\label{eq:initial-state}
	\Wo_0(x,p) &= 
    \frac{1}{2\pi h}\exp\left[-\frac{x^2+p^2}{2h}\right] = \fo_0(x,p),
\end{align}
i.e., a Gaussian state with 
$\sigma_x = h^{\frac12}=\sigma_p$.  
On the quantum side it corresponds to the position- and momentum-space wavefunctions:
\begin{align}\label{eq:initial-state-wave}
	\psi_0(x) &= \frac{1}{(2\pi h)^{\frac14}}\exp\left[-\frac{x^2}{4h}\right], \\
	\hat{\psi}_0(p) &= \frac{1}{(2\pi h)^{\frac14}}\exp\left[-\frac{p^2}{4h}\right]
\end{align}

\subsection{Hamiltonian schedule}

\begin{figure}[t]
    \centering
	\includegraphics[width=0.7\linewidth]{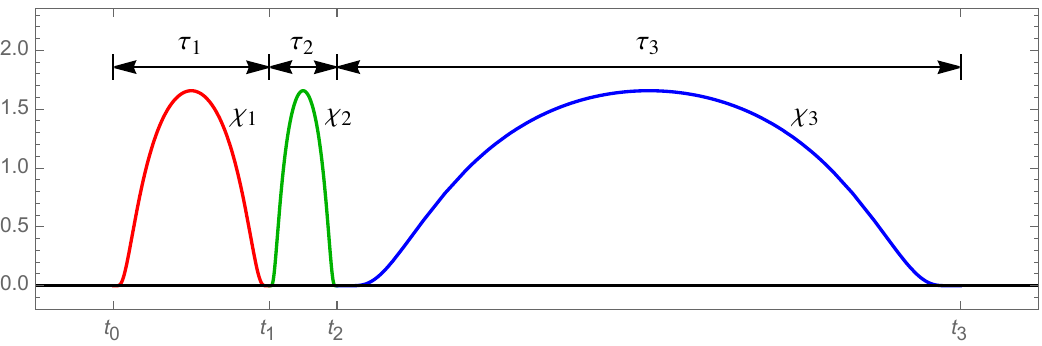}
	\caption{
        The three smooth bumps $\chi_{i=1,2,3}$ depicted respectively in red, green, and blue. They vanish outside the intervals $[t_{i-1},t_i]$ of width $\tau_i = t_i-t_{i-1}$ and satisfy $\int\! \chi_i(t) \, dt= \tau_i$.
		For the illustration we have chosen the 
		specific shared bump shape described by  \eqref{eq:chi-first}--\eqref{eq:chi-last}.
		We have also chosen the specific semiclassical parameter $h=\hbar/2=10^{-6}$ along with our standard choices $\tau_1\eqq\frac{1}{6}\log h^{-1}$, $\tau_2\eqq 1$, and $\tau_3\eqq \frac{2}{3}\log h^{-1}$.
	}
	\label{fig:bump}
\end{figure}

Consider the quantum Hamiltonian
\begin{gather}
	\label{eq:ham}
	\HQ(t) := \,\chi_1(t) \HQ_1 + \chi_2(t) \HQ_2 + \chi_3(t) \HQ_3,\\
    \nonumber\\
	\HQ_1 :=\, \frac{\hat{x}\hat{p} + \hat{p}\hat{x}}{2} =: -\HQ_3, \qquad \qquad
	\HQ_2 :=\, -\frac{1}{3} \hat{x}^3,
\end{gather}
where the $\chi_i$ are smooth bounded non-negative bump functions that turn on and off sequentially with respective durations $\tau_i$. Specifically,
\begin{align}
  t_0&=0,
  & 0\le \chi_i(t) &< \mathrm{const},\\
  t_1&=\tau_1,
  & \mathrm{supp}(\chi_i) &= [t_{i-1},t_i],\\
  t_2&=\tau_1+\tau_2,
  & \int\!\chi_i(t)\,dt &= \tau_i > 0,\\
  t_3&=\tau_1+\tau_2+\tau_3. & & 
\end{align}
Thus the evolution takes place as three discrete steps. 
For concreteness (see Figure~\ref{fig:bump}), one can imagine that the $\chi_i$ are variants of a single unit-integral bump function $\chi$,
\begin{align}
  \chi_i(t) &= \chi\!\left(\frac{t-t_{i-1}}{\tau_i}\right), \label{eq:chi-first}\\
  \chi(s) &\propto
  \begin{cases}
    \exp\left(\frac{1}{4s(s-1)}\right),& \text{if } s\in[0,1],\\
    0,& \text{otherwise},
  \end{cases}\\
  \int\!\chi(s)\,ds &= 1, \label{eq:chi-last}
\end{align}
although our results do not depend on this. The corresponding classical Hamiltonian under Wigner-Weyl is simply $\HC(t) =\chi_1(t) \HC_1 + \chi_2(t) \HC_2 + \chi_3(t) \HC_3$ with $\HC_1 = xp = -\HC_3$, $\HC_2 = -x^3/3$.

For the main calculation we will keep the step durations $\tau_i > 0$ as free variables, with the exception that we require $\tau_1 < \tfrac{1}{4}\log h^{-1}$ for technical reasons. Ultimately we will take them to be
\begin{align}
    \tau_1 \eqq \frac{1}{6}\log h^{-1}, \qquad\qquad
    \tau_2 \eqq 1, \qquad \qquad
    \tau_3 \eqq \frac{2}{3}\log h^{-1},
\end{align}
Here, and below, the notation ``$\eqq$'' denotes an equality conditional on these specific choices for the $\tau_i$.

\subsection{Conventions}

We will work in the Schr\"odinger picture, except in Section~\ref{sec:moments} where the Heisenberg picture will be advantageous.

Throughout we will slightly abuse notation by using a subscript $t$ for the general time-dependent functions (e.g., $\psi_t(x)$, $\Wo_t(x,p)$) but integer subscripts $0$, $1$, $2$, and $3$ for the specific checkpoints $t_{0,1,2,3}$ (e.g., $\psi_1(x)=\psi_{t_1}(x)$, $\Wo_2(x,p)=\Wo_{t_2}(x,p)$); this avoids nested subscripts.

\section{Pure evolution}\label{sec:pure-evolution}

First we obtain exact expressions for the pure closed-system ($D=0$) evolution. For the closed systems we will use a circle accent to denote the quantum state $\rhoc_t$ and the classical state $\fc_t$.  (Later, as an auxiliary, we will need to introduce a hybrid trajectory that is sometimes open and sometimes closed for which we will use a tilde: $\rhoh_t, \fh_t$.)  Only the closed quantum system will be pure, so we do not bother to put an accent on the wavefunction $\psi$, i.e., $\rhoc_t = |\psi_t\rangle\langle\psi_t|$.  We also use $\psi_t(x) = \langle x | \psi_t\rangle$ to denote the position-space wavefunction and $\hat\psi_t(p) = \langle p | \psi_t\rangle = (2\pi\hbar)^{-\frac12}\int\! \diff x \,\psi_t(x) e^{-ixp/\hbar}$ for the momentum-space wavefunction.  Our Fourier convention is symmetric: $\psi_t(x) = (2\pi\hbar)^{-\frac12}\int\! \diff p\, \hat\psi_t(p) e^{ixp/\hbar}$.

\subsection{Classical closed-system evolution}

The classical equations of motion under $\HC_1$ are 
\begin{align}
	\dot{x} &= \, \partial_p H_1 = x,\\
	\dot{p} &= -\partial_x H_1 = -p,
\end{align}
so $x(t) = x_0 e^{t}$, $p(t) = p_0 e^{-t}$. After the first classical step of duration $\tau_1$, the initial state \eqref{eq:initial-state} is evolved by the closed-system dynamics to
\begin{align}
	\fc_1(x,p) &= \fc_0 (x e^{-\tau_1}, p e^{\tau_1}) \\
	&= \frac{1}{2\pi h}\exp\!\left[-\frac{x^2}{2 h e^{2\tau_1}}-\frac{p^2}{2 h e^{-2\tau_1}}\right],
\end{align}
i.e., an (anti-squeezed) Gaussian wavepacket with $\Delta x_1 = h^{\frac12} e^{\tau_1} \ \eqq \ h^{\frac13}$, $\Delta p_1 = h^{\frac12} e^{-\tau_1} \ \eqq \ h^{\frac23}$.

The classical equations of motion under $\HC_2$ are 
\begin{align}
	\dot{x} &= \, \partial_p H_2 =0,\\
	\dot{p} &= -\partial_x H_2 =  x^2,
\end{align}
so $x(t) = x_1$, $p(t) = p_1 +  (t-t_1) x_1^2$. After the second classical step of duration $\tau_2$ we have 
\begin{align}
	\fc_2 (x,p) &= \fc_1(x,p-\tau_2 x^2)  \\
	&= \frac{1}{2\pi h}\exp\!\left[-\frac{x^2}{2 h e^{2\tau_1}}-\frac{(p-\tau_2 x^2)^2}{2 h e^{-2\tau_1}}\right] ,
\end{align}
and marginalizing over $x$ we have the classical momentum distribution
\begin{align}
	\cmc_2 (p) &=\int_{-\infty}^\infty 	\fc_2 (x,p)\, \diff x  \\
    &= \frac{1}{2\pi h}\exp\!\left[-\frac{p^2}{2 h e^{-2\tau_1}}\right]
    \int_{-\infty}^\infty 
    \exp\!\left[-\frac{x^2\!\left(e^{-4\tau_1}-2p \tau_2\right)+\tau_2^2 x^4}{2 h e^{-2\tau_1}}\right] \diff x \\
    &= \frac{1}{2 \sqrt{\pi \tau_2}\,\sqrt{h e^{-2\tau_1}}}\,
    \exp\!\left[
    -\frac{1}{2}\!\left(\frac{p}{\sqrt{h e^{-2\tau_1}}}\right)^{\!2}
    +\frac{1}{4}\!\left(\frac{p }{\sqrt{h e^{-2\tau_1}}}-\frac{1}{2 \tau_2}\right)^{\!2}
    \right]\,
    \pcf_{-\frac12}\!\left(\frac{1}{2 \tau_2} - \frac{p }{\sqrt{h e^{-2\tau_1}}}\right).
\end{align}
To get the third line we have used a substitution $s:=x^2$ and identified the integral as an example of the parabolic cylinder function
\begin{align}
\pcf_{\ell}(z) = \frac{e^{-\tfrac{z^{2}}{4}}}{\Gamma(-\ell)}
\int_{0}^{\infty} e^{-z s - s^{2}/2} \, s^{-\ell-1} \, ds \qquad  \qquad (\mathrm{Re}\, \ell <0 )
\end{align}
See Eq.~(2) in Subsection~9.241 (of Section~9.24) in Gradshteyn \& Ryzhik \cite{gradstejn2009table}.
($\Gamma$ is the Gamma function, and we use $\Gamma(\tfrac12)=\pi^{\frac12}$.)

The classical equations of motion under $\HC_3$ are 
\begin{align}
	\dot{x} &= \, \partial_p H_3 = - x,\\
	\dot{p} &= -\partial_x H_3 =  p,
\end{align}
so $x(t) = x_2 e^{-(t-\tau_2)}$, $p(t) = p_2 e^{t-\tau_2}$. After the third (last) classical step of duration $\tau_3$ we have the final momentum distribution
\begin{align}\label{eq:cmc3}
	\cmc_3(p) &= e^{-\tau_3}\, \cmc_2 (p\, e^{-\tau_3})  \\
    &= \frac{1}{2 \sqrt{\pi \tau_2}}\frac{1}{h^{\frac12}e^{\tau_3-\tau_1}}\,
    \exp\!\left[
    -\frac{1}{2}\!\left(\frac{p}{ h^{\frac12}e^{\tau_3-\tau_1}}\right)^{\!2}
    +\frac{1}{4}\!\left(\frac{p}{ h^{\frac12}e^{\tau_3-\tau_1}}-\frac{1}{2 \tau_2}\right)^{\!2}
    \right]\,
    \pcf_{-\frac12}\!\left(\frac{1}{2 \tau_2} - \frac{p}{ h^{\frac12}e^{\tau_3-\tau_1}}\right)\\
	&\eqq \frac{1}{2 \sqrt{\pi}}\,
	\exp\!\left[-\frac{p^2}{2}+\frac{1}{4}\left(p -\frac12 \right)^2\right]\,
	\pcf_{-\frac12}\!\left(\frac12  - p\right) ,
\end{align}
which, for our choices of $\tau_1$ and $\tau_3$, is independent of $h$.  (Note that $h$-independence requires $\tau_3-\tau_1 = \frac{1}{2}\log h^{-1}$, but that $\tau_1$ and $\tau_3$ are not individually fixed by this requirement.)

\subsection{Quantum closed-system evolution}

The first quantum step stretches (anti-squeezes) in the $x$ direction by a factor $e^{\tau_1}$ so
\begin{align}
	\psi_1 (x) &= 	e^{-\tau_1/2} \, \psi_0 (x e^{-\tau_1})\\
	&=  (2\pi h e^{2\tau_1})^{-\frac14}\exp\left[-\frac{x^2}{4h e^{2\tau_1}}\right] 
\end{align}
The second quantum step is diagonal in the $x$ basis with $\HC_2 = -x^3/3$ and duration $\tau_2$, so
\begin{align}
	\psi_2 (x) &= 	\exp\!\left(-\frac{i \tau_2 H_2}{\hbar}\right) \psi_1 (x)\\
	&=  (2\pi h e^{2\tau_1})^{-\frac14}\exp\!\left[-\frac{x^2}{4h e^{2\tau_1}}+ i\,\frac{\tau_2 x^3}{6h}\right] 
\end{align}
(where we recall $\hbar=2h$).
In momentum space this is 
\begin{align}
	\hat{\psi}_2 (p) &= \frac{1}{(2\pi\hbar)^{\frac12}} \int_{-\infty}^\infty e^{-i x p /\hbar} \psi_2 (x) \, \diff x\\
    &= \frac{1}{2^{\frac54}\pi^{\frac34}h^{\frac34}e^{\tau_1/2} }  \int_{-\infty}^\infty \exp\left[-\frac{x^2}{4 h e^{2\tau_1}}-\frac{i p x}{2 h}+\frac{i \tau_2  x^3}{6 h}\right] \, \diff x\\
    &= \frac{2^{\frac{1}{12}} \pi^{\frac14} }{ \tau_2^{\frac13}h^{\frac{5}{12}}e^{\tau_1/2}} 
    \exp\!\left[\frac{\tau_2^{-1 }e^{-4\tau_1}-6 p}{24 \tau_2 h e^{2\tau_1}}\right]
	\mathrm{Ai}\left(\frac{\tau_2^{-1 }e^{-4\tau_1}-4 p}{2^{\frac83} \tau_2^{\frac13}h^{\frac23}}\right),
\end{align}
where the integral is evaluated using the substitution
$x=a (2h/\tau_2)^{\frac13}-i e^{-2\tau_1}/(2\tau_2)$ 
and the (distributional) integral identity for the Airy function  $\mathrm{Ai}(z)$, 
\begin{align}
	\int_{-\infty}^\infty e^{i(a^3/3+za)}  \, da = 2\pi \mathrm{Ai}(z).
\end{align}
The third quantum step stretches (anti-squeezes) in the $p$ direction by a factor $e^{\tau_3}$, hence
\begin{align}
	\hat{\psi}_3 (p) &= e^{-\tau_3/2}\, \hat{\psi}_2 (p\, e^{-\tau_3})
\end{align}
and 
\begin{align}
	\qmc_3(p) & = |\hat{\psi}_3 (p)|^2 \\
	&= e^{-\tau_3} |\hat{\psi}_2 (p e^{-\tau_3})|^2\\
    &= \frac{2^{\frac{1}{6}} \pi^{\frac12} }{ \tau_2^{\frac23}  h^{\frac56}e^{\tau_1+\tau_3}} 
    \exp\!\left[\frac{\tau_2^{-1 }e^{\tau_3-4\tau_1}-6 p }{12 \tau_2 h e^{2\tau_1+\tau_3}}\right]
	\mathrm{Ai}\left(\frac{\tau_2^{-1 }e^{\tau_3-4\tau_1}-4 p}{2^{\frac83} \tau_2^{\frac13}h^{\frac23}e^{\tau_3}}\right)^2,\\
    &\eqq 2^{\frac{1}{6}} \pi^{\frac12}  
    \exp\!\left[\frac{1-6 p }{12 }\right]
	\mathrm{Ai}\left(\frac{1-4 p}{2^{\frac83} }\right)^2.\label{eq:q3}
\end{align}
The choices $\tau_1 \eqq \tfrac{1}{6}\log h^{-1}$ and $\tau_3 \eqq \tfrac{2}{3}\log h^{-1}$ eliminate the $h$ dependence (and indeed up to an additive $h$-independent constant they are the only ones that do so). 

\begin{figure}[ht]
	\includegraphics[width=0.95\linewidth]{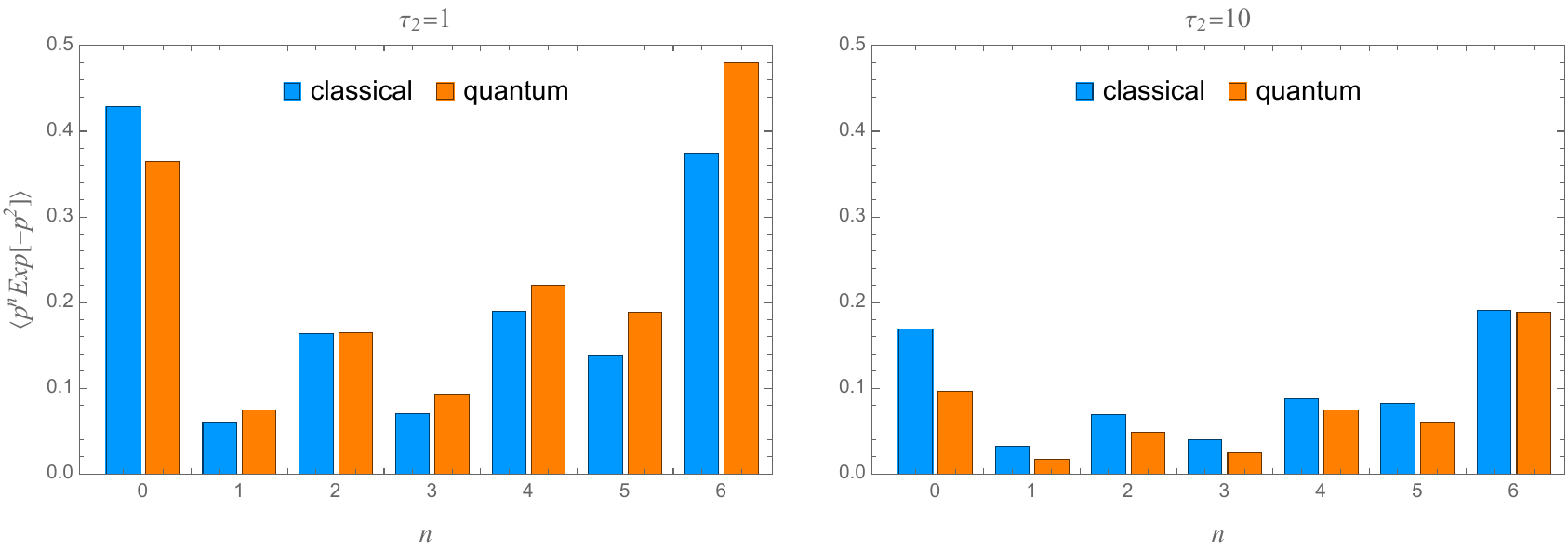}
	\caption{
		The expectation values $\langle \gp_n\rangle_{\fc_3}$ and $\langle \hat{\gp}_n\rangle_{\psi_3}$ of the $n$-indexed bounded smooth observable $\gp_n :=  p^n \exp[-p^2]$ with respect to the classical (blue) closed-system momentum distribution $\cmc_3(p)$ vs.\ the quantum (orange)  closed-system momentum distribution $\qmc_3(p) = |\hat{\psi}_3 (p)|^2$. We choose $\tau_1=\tfrac{1}{6}\log h^{-1}$ and $\tau_3=\tfrac{2}{3}\log h^{-1}$, ensuring the functions are independent of $h$, and compare $\tau_2 = 1$ (left) with $\tau_2 = 10$ (right).  See Fig.~\ref{fig:p-distribution} for the distributions themselves.
	}
	\label{fig:p-observables}
\end{figure}

In Fig.~\ref{fig:p-distribution}, we compare the classical and quantum momentum distributions for negligible diffusion. In Fig.~\ref{fig:p-observables} we do the same for smooth observable expectation values.

\subsection{Second and fourth central moments at checkpoints}
\label{sec:moments}

(We will use $x$ and $p$ both as parameters and as random variables, using context to distinguish.)

In this section we calculate the second and fourth central moments of $x$ and $p$ during the quantum and classical evolution.  This will help us compare the open and closed evolution later in Section~\ref{sec:diffusion-error}. The calculation of moments is straightforward because, in the classical case, the variables are initially independent Gaussians and evolve as
\begin{align}
	x &\mapsto e^{\tau_1} x,&  p &\mapsto e^{-\tau_1} p, \\
	x &\mapsto x,& p &\mapsto p+\tau_2 \, x^2, \\
	x &\mapsto e^{-\tau_3} x,& p &\mapsto e^{\tau_3} p, 
\end{align}
during the three steps generated respectively by $\HC_1, \HC_2, \HC_3$. (Thus the classical state is always a ``conditional Gaussian'': it takes the form $\fc(x,p) = \gauss_{\sigma_x}(x)\gauss_{\sigma_p}(p-r x^2)$ for all times, where $\gauss_\sigma$ is the centered Gaussian with variance $\sigma^2$.)

Furthermore, in the quantum case, the Heisenberg-picture versions of these variables evolve the same way. For example, under $\HC_2$ the quantum variables transform as $\hat{p}_2 = \hat{p}_1+\tau_2 \, \hat{x}_1^2$. To see this recall that $[g(x),p]=i\hbar g'(x)$. 
As we will show, the $x$-distributions for the quantum and classical systems are identical and the $p$-distributions, although not identical, have the same first, second, and fourth central moments.

For clarity we compute moments of $x_i,p_i$ at the checkpoint $i\in\{0,1,2,3\}$ that are partitioned by the steps.  The interpolation for intermediate times (during the $\chi_i$ bumps) is obvious.  

\subsubsection{Classical case}

Denote  
the expectations $\bar{x}_i :=  \langle x_i\rangle$ and $\bar{p}_i :=  \langle p_i\rangle$.
The only non-zero means are $\bar{p}_2$ and $\bar{p}_3$:
\begin{alignat}{3}\label{moment-start}
\bar{x}_0&=0, 
\qquad\qquad\qquad\qquad 
&\bar{p}_0&=0, 
& &\\
\bar{x}_1&=0,
\qquad\qquad\qquad\qquad 
&\bar{p}_1&=0, 
& &\\
\bar{x}_2&=0,
\qquad\qquad\qquad\qquad &\bar{p}_2&=\tau_2\,h\,e^{2\tau_1} 
& &\eqq h^{\frac23},\\
\bar{x}_3&=0,
\qquad\qquad\qquad\qquad &\bar{p}_3&=\tau_2\,h\,e^{2\tau_1+\tau_3} 
& &\eqq 1 .
\end{alignat}
At $i=0$ (initial Gaussian; $\Delta x=\Delta p=h^{\frac12}$):
\begin{align}
	\langle (x_0-\bar{x}_0)^2\rangle &= h,\\
	\langle (x_0-\bar{x}_0)^4\rangle &= 3 h^{2},\\
	\langle (p_0-\bar{p}_0)^2\rangle &= h,\\
	\langle (p_0-\bar{p}_0)^4\rangle &= 3 h^{2}.
\end{align}
After step 1 ($\HC_1$ anti-squeezing; $\Delta x=h^{\frac12}e^{\tau_1}$, $\Delta p=h^{\frac12}e^{-\tau_1}$):
\begin{alignat}{2}
	\langle (x_1-\bar{x}_1)^2\rangle 
    &= h e^{2\tau_1} 
    & &\eqq \ h^{\frac23},\\
	\langle (x_1-\bar{x}_1)^4\rangle 
    &= 3 h^{2} e^{4\tau_1} 
    & &\eqq \ 3 h^{\frac43},\\
	\langle (p_1-\bar{p}_1)^2\rangle 
    &= h e^{-2\tau_1} 
    & &\eqq \ h^{\frac43},\\
	\langle (p_1-\bar{p}_1)^4\rangle 
    &= 3 h^{2} e^{-4\tau_1}
    & &\eqq \ 3 h^{\frac83}.
\end{alignat}
After step 2 ($\HC_2$ kick; $x\mapsto x$, $p\mapsto p+\tau_2 x^2$):
\begin{alignat}{2}
	\langle (x_2-\bar{x}_2)^2\rangle 
    &= h e^{2\tau_1} 
    & &\eqq \ h^{\frac23},
    \\
	\langle (x_2-\bar{x}_2)^4\rangle 
    &= 3 h^{2} e^{4\tau_1}
    & &\eqq \ 3 h^{\frac43},
    \\
	\langle (p_2-\bar{p}_2)^2\rangle 
    &= 
    h e^{-2\tau_1}(1+ 2 \tau_2^2 h e^{6\tau_1})
    & &\eqq\ 3 h^{\frac43}, \label{eq:tricky1c}
    \\
	\langle (p_2-\bar{p}_2)^4\rangle 
    &= 
    3h^2 e^{-4\tau_1}(
    1
    +4\tau_2^2 h e^{6\tau_1}
    +20\tau_2^4 h^2 e^{12\tau_1}
    )
    & &\eqq\ 75 h^{\frac83}. \label{eq:tricky2c}
\end{alignat}
After step 3 ($\HC_3$ squeeze; $x\mapsto e^{-\tau_3}x$, $p\mapsto e^{\tau_3}p$):
\begin{alignat}{2}
	\langle (x_3-\bar{x}_3)^2\rangle 
    &= h \, e^{2(\tau_1-\tau_3)}
    & &\eqq\ h^{2},\\
	\langle (x_3-\bar{x}_3)^4\rangle 
    &= 3 h^{2} e^{4(\tau_1-\tau_3)}
    & &\eqq\ 3 h^{4},\\
	\langle (p_3-\bar{p}_3)^2\rangle 
    &= 
    h e^{2(\tau_3-\tau_1)}(1+ 2 \tau_2^2 h e^{6\tau_1})
    & &\eqq\ 
    3,\\
    \label{moment-end}
	\langle (p_3-\bar{p}_3)^4\rangle 
    &= 
    3h^2 e^{4(\tau_3-\tau_1)}(
    1
    +4\tau_2^2 h e^{6\tau_1}
    +20\tau_2^4 h^2 e^{12\tau_1}
    )
    & &\eqq\ 
    75.
\end{alignat}

Most of these are trivial. Let us explain \eqref{eq:tricky1c} and \eqref{eq:tricky2c}. Let $\sigma_x:= h^{\frac12} e^{\tau_1} \eqq h^{\frac13}$ and $\sigma_p:=h/\sigma_x=h^{\frac12} e^{-\tau_1}\eqq h^{\frac23}$. After step 1,
$p_1\sim\mathcal{N}(0,\sigma_p^2)$, $x_1\sim\mathcal{N}(0,\sigma_x^2)$, and they are independent. Furthermore, $x_2 = x_1$ and $p_2 = p_1+\tau_2 x_1^2$.
Write $z\sim\mathcal{N}(0,1)$ so $x_2=\sigma_x z$ and set
$A:=p_1$, $B:=x_1^2-\sigma_x^2=x_2^2-\sigma_x^2=\sigma_x^2(z^2-1)$.
Then $p_2-\bar p_2=A+\tau_2 B$.

Denote the $n$th moment and cumulant by $\mu_n$ and $\kappa_n$, respectively. Cumulants add for independent sums, i.e., $\kappa_n(A+B) = \kappa_n(A)+ \kappa_n(B)$ for independent $A$ and $B$.
They also obey $\kappa_r(cY)=c^r\kappa_r(Y)$.
For Gaussian $A$ we have $\kappa_2(A)=\sigma_p^2$, $\kappa_4(A)=0$.
For $z^2$ (a chi-squared distribution with 1 degree of freedom): $\kappa_2(z^2)=2$, $\kappa_4(z^2)=48$.
Hence $\kappa_2(B)=2\sigma_x^4$ and $\kappa_4(B)=48\sigma_x^8$, so
\begin{align}
\kappa_2(p_2-\bar p_2)&=\kappa_2(A)+\tau_2^2 \kappa_2(B) = \sigma_p^2+2\tau_2^2\sigma_x^4,\\
\kappa_4(p_2-\bar p_2)&=\kappa_4(A)+\tau_2^4 \kappa_4(B) = 48\tau_2^4\sigma_x^8.
\end{align}
Using the moment-cumulant relations $\mu_2=\kappa_2$ and $\mu_4=\kappa_4+3\kappa_2^2$, we have
\begin{align}
\langle (p_2-\bar p_2)^2\rangle
&=\sigma_p^2+2\tau_2^2\sigma_x^4\\
&= h e^{-2\tau_1} +2\tau_2^2 h^2 e^{4\tau_1}
\end{align}
and
\begin{align}
\langle (p_2-\bar p_2)^4\rangle
&=48\tau_2^4\sigma_x^8+3(\sigma_p^2+2\tau_2^2\sigma_x^4)^2\\
&=48 \tau_2^4 h^4 e^{8\tau_1}+3(h e^{-2\tau_1}+2h^2 \tau_2^2e^{4\tau_1})^2
\end{align}
which are \eqref{eq:tricky1c} and \eqref{eq:tricky2c}.

\subsubsection{Quantum case}

Now we show that the quantum system has the exact same values for the moments discussed above in the classical case, \eqref{moment-start}--\eqref{moment-end}.

For the initial state ($t_0=0$) and after the first time step ($t_1=\tau_1$), the state is Gaussian with a diagonal covariance matrix, so the expectations of all powers of $x$ are identical in the quantum and classical cases, and likewise for $p$.  Furthermore, the Hamiltonian $\HC_2(x,p)=-x^3/3$ of the second step is diagonal in $x$, so it does not change the $x$-distribution, and the third step just squeezes in $x$ (i.e., $\hat{x}_3 = \hat{x}_2 e^{-\tau_3}$), so the quantum and classical $x$-distributions will remain the same for all times. 

For $p$, we will only address the first, second and fourth central moments.\footnote{The other moments of $p$ will in general be \emph{different} for quantum vs.\ classical, as indeed we need for our counterexample.} It will suffice to show that these moments are the same for the quantum and classical states after step 2 because the third step just stretches in $p$ (i.e., $\hat{p}_3 = \hat{p}_2 e^{\tau_3}$). 

Let $\Wc(x,p,t)$ be the Wigner function. In general, the Moyal equation (i.e., quantum Liouville equation) for the dynamics of the Wigner function contains arbitrarily high powers of $x$- and $p$-derivatives of $\Wc$ and of the Hamiltonian $H$. However, for the specific Hamiltonian $\HC_2(x,p) =  -x^3/3$ during the second step, which has no $p$-dependence and only cubic $x$-dependence, the Moyal equation truncates at third order, simplifying to
\begin{align}
\partial_t \Wc
&= \partial_x H_2\partial_p \Wc
-\frac{\hbar^2}{24}(\partial_x^3 H_2)(\partial_p^3\Wc)\\
&= -x^2 \partial_p \Wc
+\frac{h^2}{3} \partial_p^3 \Wc
\end{align}
(recalling $h=\hbar/2$). 
This differs from the classical Liouville equation for the classical distribution $\fc$ only by the second term:
\begin{align}
	\partial_t \fc
	&= \{\!\{H_2, \fc\}\!\}_{\rm PB} \\
	&= -x^2 \partial_p \fc
\end{align}
Let us use the hat to denote the Fourier transform with respect to $p$:
\begin{align}
\widehat W(x,k,t):=\int e^{ikp}\Wc(x,p,t)\,\diff p.\\
\widehat f(x,k,t):=\int e^{ikp}\fc(x,p,t)\,\diff p.
\end{align}
Then $\partial_p \mapsto ik$ and $\partial_p^{3}\mapsto (ik)^3=-ik^3$, so
\begin{align}
\partial_t \widehat W
&= -ik x^2 \widehat W 
- i\frac{h^2}{3}k^3 \widehat W,\\
\partial_t \widehat f
&= -ik x^2 \widehat f
\end{align}
Thus, since the quantum and classical distributions over phase space are identical after step 1, $\Wc(x,p,t_1)=\fc(x,p,t_1)$, we have for any $t\in[t_1,t_2]$ that
\begin{align}
\widehat W(x,k,t)
= e^{-i h^2 (t-t_1)k^3/3}\;\widehat f(x,k,t).
\end{align}
The respective cumulant generating functions for the $p$ distribution are
\begin{align}
	K_{\rm qm}(k,t) & :=  \log\, \langle e^{i k \hat{p}}\rangle_{\rm qm}(t)= \log \int \widehat W(x,k,t)\,\diff x\\
	K_{\rm cl}(k,t) & :=  \log \, \langle e^{i k p}\rangle_{\rm cl}(t)= \log \int \widehat f(x,k,t)\,\diff x
\end{align}
so they are related by
\begin{align}
K_{\rm qm}(k,t)=K_{\rm cl}(k,t)-i\frac{h^2 (t-t_1)}{3}k^3.
\end{align}
For any random variable, the $n$th cumulant $\kappa_n$ is
\[
\kappa_n(t)=\frac{1}{i^n}\frac{d^n}{dk^n}K(k,t)\Big|_{k=0}.
\]
Since the difference is proportional to $k^3$, it contributes only to the third cumulant $\kappa_3$. 

The second and fourth central moments $\mu_2 := \langle (p-\bar p)^2\rangle$ and $\mu_4 := \langle (p-\bar p)^4\rangle$ can be written in terms of the cumulants as $\mu_2=\kappa_2$ and $\mu_4=\kappa_4+3\kappa_2^2$. These depend only on $\kappa_2$ and $\kappa_4$, both of which are identical in quantum and classical cases. As mentioned, the simple stretching of step 3 ensures there is no quantum-classical discrepancy generated then either.

\section{Bounding the diffusion error}\label{sec:diffusion-error}

Now we consider the effects of the diffusion of strength $D$.

\subsection{Quantum diffusion error}

Let $\rhoc_t = |\psi_t\rangle\langle\psi_t|$ be the solution to the quantum closed-system dynamics
\begin{align}
	\frac{\partial\rhoc}{\partial t} &= \LLQ_H[\rhoc] = -i\hbar^{-1}[\HQ,\rhoc]\\
	&= -i (2h)^{-1}[\HQ,\rhoc]
\end{align}
as described above, where $H=H(t)$ is the three-bump Hamiltonian given by \eqref{eq:ham}.  We will compare it to the open-system trajectory $\rhoo_t$, starting from the initial pure Gaussian state $\rhoo_0 = |\psi_0\rangle\langle\psi_0|$, \eqref{eq:initial-state-wave}, but which instead solves
\begin{align}
	\frac{\partial\rhoo}{\partial t} = \LLQ[\rhoo] = \LLQ_H[\rhoo]+\LLQ_D[\rhoo].
\end{align}
Here, the diffusion Lindbladian is 
\begin{align}
	\LLQ_D[\rhoo] &:= -\frac{D}{2\hbar^{2}} ( [\hat x,[\hat x,\rhoo]] + [\hat p,[\hat p,\rhoo]] )\\
	&= -\frac{D}{8h^{2}} ( [\hat x,[\hat x,\rhoo]] + [\hat p,[\hat p,\rhoo]] ),
\end{align}
which corresponds, in the Wigner representation, to the differential operator
\begin{align}\label{eq:diff-superop}
	{\LLC}_D[\Wo] = \frac{1}{2} D (\partial^2_x \Wo + \partial^2_p \Wo).
\end{align}
We want to show that introducing diffusion produces only an asymptotically negligible change to the final momentum distribution $\qmc_3$, i.e., that $\|\qmo_3-\qmc_3\|_{L^1}=o(h^0)$.

\subsubsection{Duhamel bound on growth of overall state error \texorpdfstring{$\|\rhoo_t-\rhoc_t\|_{\mathrm{Tr}}$}{}}

By Duhamel, 
\begin{align}
	\rhoo_t-\rhoc_t = \int_0^t \! ds \, e^{(t-s) \LLQ} \left(\LLQ-\LLQ_H\right)[\rhoc_s]
\end{align}
we have 
\begin{align}
	\|\rhoo_t-\rhoc_t\|_{\mathrm{Tr}} &\le \int_0^t \! ds \, \| \LLQ_D[\rhoc_s] \|_{\mathrm{Tr}} \\
	\label{eq:duhamel-bound}
	&\le \frac{tD}{8h^2} \sup_{s\in[0,t]} \left\| [x,[x,\rhoc_s]] + [p,[p,\rhoc_s]] \right\|_{\mathrm{Tr}}.
\end{align}
Let us focus for the moment on a particular time $s$ inside the supremum.  
For the $x$ double commutator,
\begin{align}
	\| [x,[x,\eta]] \|_{\mathrm{Tr}} = \| x^2\eta - 2 x\eta x + \eta x^2 \|_{\mathrm{Tr}} 
	&\le 2\left( \|x^2 \eta\|_{\mathrm{Tr}} + \|x \eta x\|_{\mathrm{Tr}}\right).
\end{align}
We bound the first term with H\"older's inequality, $\|AB\|_{\mathrm{Tr}} \le \|A\|_{\mathrm{HS}} \|B\|_{\mathrm{HS}}$ (where $\|\cdot\|_{\mathrm{HS}}$ denotes the Hilbert-Schmidt norm),
\begin{align}
	\|x^2 \eta\|_{\mathrm{Tr}} \le \| x^2 \sqrt{\eta} \|_2 \| \sqrt{\eta} \|_{\mathrm{HS}} = \langle x^4\rangle^{\frac12},
\end{align}
and the second term using $x \eta x \ge 0$,
\begin{align}
	\|x \eta x\|_{\mathrm{Tr}} = \Trace(x \eta x) = \langle x^2\rangle,
\end{align}
so that 
\begin{align}
	\| [x,[x,\eta]] \|_{\mathrm{Tr}} \le 2\left(\langle x^2\rangle_\eta + \langle x^4\rangle^{\frac12}_\eta\right).
\end{align}
Using this and the corresponding bound for $\left\| [p,[p,\eta]] \right\|_{\mathrm{Tr}}$ gives 
\begin{align}
\left\| [x,[x,\rhoc_s]] + [p,[p,\rhoc_s]] \right\|_{\mathrm{Tr}} 
\le 2\left(\langle x^2\rangle_{\psi_s} + \langle x^4\rangle^{\frac12}_{\psi_s} + \langle p^2\rangle_{\psi_s} + \langle p^4\rangle_{\psi_s}^{\frac12}\right)
\end{align}
More generally, we can get this for central moments by repeating the last few steps using $\tilde{x} := x-\bar{x} = x - \langle x\rangle_{\psi_s}$ and $\tilde{p} := p-\bar{p} = p - \langle p\rangle_{\psi_s}$ to get 
\begin{align}
\left\| [x,[x,\rhoc_s]] + [p,[p,\rhoc_s]] \right\|_{\mathrm{Tr}} 
&=\left\| [\tilde x,[\tilde x,\rhoc_s]] + [\tilde p,[\tilde p,\rhoc_s]] \right\|_{\mathrm{Tr}}\\
&\le 2\left(\langle (x-\bar{x})^2\rangle_{\psi_s} + \langle (x-\bar{x})^4\rangle^{\frac12}_{\psi_s} + \langle (p-\bar{p})^2\rangle_{\psi_s} + \langle (p-\bar{p})^4\rangle_{\psi_s}^{\frac12}\right)
\end{align}
Inserting this into \eqref{eq:duhamel-bound} gives
\begin{align}\label{eq:kick-sup}
    \|\rhoo_t-\rhoc_t\|_{\mathrm{Tr}} 
    &\le \frac{tD}{4h^2} \sup_{s\in[0,t]} \left[ \langle (x-\bar{x})^2\rangle_{\psi_s} + \langle (x-\bar{x})^4\rangle^{\frac12}_{\psi_s} + \langle (p-\bar{p})^2\rangle_{\psi_s} + \langle (p-\bar{p})^4\rangle_{\psi_s}^{\frac12}\right]
\end{align}

However, the effect of the $x$-diffusion (i.e., $p$-decoherence) on the quantum state is actually large (in trace norm) during the third step when the state becomes highly squeezed in $x$ (stretched in $p$)  such that $\langle p^2\rangle_{\psi_3} \sim O(h^0)$. We would have to take $D=o(h^2)$. So we cannot take $t$ into the third step.

But that's okay because we are only interested in measuring the distribution of $p$, which is only affected by the $p$-diffusion (i.e., $x$-decoherence).  The plan is to bound the small error introduced by diffusion on the overall state -- and thus to the $p$ distribution -- up through the end of the second step, and then separately bound the additional error to the $p$ distribution due to diffusion in the third step.

More specifically: We will introduce an intermediate (``hybrid'') quantum trajectory $\rhoh_t$ that is equal to the closed-system evolution $\rhoc_t$ up to the end of the second step  ($\rhoh_2=\rhoc_2$) and which then experiences the diffusion during the third step (like $\rhoo$).  We will bound $\|\qmo_3 - \qmc_3\|_{L^1}$ by starting with the triangle inequality with the intermediate trajectory
\begin{align}\label{eq:triangle}
	\|\qmo_3-\qmc_3\|_{L^1}  &\le  \|\qmo_3-\qmh_3\|_{L^1} + \|\qmh_3-\qmc_3\|_{L^1} .
\end{align}

\subsubsection{Bounding overall state error \texorpdfstring{$\|\qmo_3-\qmh_3\|_{L^1} 
		\le  \|\rhoo_2-\rhoc_2\|_{\mathrm{Tr}}$}{} through end of step 2}

For the first term on the r.h.s.\ of \eqref{eq:triangle}, the open and hybrid trajectories both evolve during the third step according to the same Lindbladian $\LLQ$ so 
\begin{align}
	 \|\qmo_3-\qmh_3\|_{L^1}  & \le \|\rhoo_3-\rhoh_3\|_{\mathrm{Tr}} \\
	& = \left\|e^{\tau_3 \LLQ}( \rhoo_2-\rhoh_2)\right\|_{\mathrm{Tr}} \\
	&\le \|\rhoo_2-\rhoh_2\|_{\mathrm{Tr}} \label{eq:cp-contract-q}\\
	& = \|\rhoo_2-\rhoc_2\|_{\mathrm{Tr}} \label{eq:rhoh2-eq-rhoc2}\\
	 &\le \frac{(\tau_1 + \tau_2)D}{4h^2}
  \left[h e^{2\tau_1} +\sqrt{3} h e^{2\tau_1} 
  + h (1+ 2 \tau_2^2 h e^{6\tau_1})
  +  h \sqrt{
    3
    +12\tau_2^2 h e^{6\tau_1}
    +60\tau_2^4 h^2 e^{12\tau_1}
    }\right]
    \label{eq:insert-moments}\\
  &\le \frac{(\tau_1 + \tau_2)D}{4h e^{-2\tau_1}}
  \left[ 1 +\sqrt{3}  
  +  (e^{-2\tau_1} + 2 \tau_2^2 h e^{4\tau_1})
  +     \sqrt{
  	3e^{-4\tau_1}
  	+12\tau_2^2 h e^{2\tau_1}
  	+60\tau_2^4 h^2 e^{8\tau_1}
  }\right]
  \\
  &< 
  \frac{(\tau_1 + \tau_2)D}{4h e^{-2\tau_1}}
  \left[1 +\sqrt{3}  
  +  (1+ 2 \tau_2^2 )
  +  \sqrt{
    3
    +12\tau_2^2 
    +60\tau_2^4 
    }\right]
    \label{eq:invoke-assump7}
  \\
  &= 
  C_1\frac{(\tau_1 + \tau_2)D}{h e^{-2\tau_1}}
  \label{eq:term4}
\end{align}
with constant 
\begin{align}
    C_1 := \frac{1}{4}\left[1 +\sqrt{3}  
  +  (1+ 2 \tau_2^2 )
  +  \sqrt{
    3
    +12\tau_2^2 
    +60\tau_2^4 
    }\right]
    \eqq 1+\frac{3\sqrt{3}}{2}
    \approx 3.598 .
\end{align}
In \eqref{eq:cp-contract-q}, we use the fact that $e^{\tau_3 \LLQ}$ is a CP map which cannot increase trace norm. In \eqref{eq:rhoh2-eq-rhoc2} we use $\rhoh_2=\rhoc_2$. To get \eqref{eq:insert-moments} we have used \eqref{eq:kick-sup} with $t=t_2=\tau_1 + \tau_2$ and the central moment bounds proven in Section~\ref{sec:moments}.  
Finally in \eqref{eq:invoke-assump7} we invoke $h\le 1$ and our technical assumption that $\tau_1 < \tfrac{1}{4}\log h^{-1}$.

\subsubsection{Bounding momentum error \texorpdfstring{$\|\qmh_3-\qmc_3\|_{L^1}$}{} during step 3}

The second term on the r.h.s.\ of \eqref{eq:triangle}, $\|\qmh_3-\qmc_3\|_{L^1}$, can be bounded by observing that the dynamics for the Wigner function during the third step decompose into terms involving only $x$ or only $p$:
\begin{align}
  \partial_t \Wc &= \LLC^{(x)}_{3,H}[\Wc]+\LLC^{(p)}_{3,H}[\Wc],\\
  \partial_t \Wh &= (\LLC^{(x)}_{3,H}+\LLC^{(x)}_D)[\Wh]+(\LLC^{(p)}_{3,H}+\LLC^{(p)}_D)[\Wh],
\end{align}
where
\begin{align}
	\LLC^{(x)}_{D} &= \frac{1}{2} D \partial_x^2,&  \LLC^{(p)}_{D} &=  \frac{1}{2} D \partial_p^2,\\
	\LLC^{(x)}_{3,H} &=  x \partial_x , &  \LLC^{(p)}_{3,H} &= - p \partial_p .
\end{align}
Thus the reduced dynamics for the momentum distributions $\qmh_t(p) := \langle p |\rhoh_t |p \rangle = \int\! \diff x\, \Wh_t(x,p)$ and $\qmc_t(p) := |\hat{\psi}_t (p)|^2 = \int\! \diff x\, \Wc_t(x,p)$ are
\begin{align}
    \partial_t \qmc &= \int\! \diff x\, \partial_t \Wc 
    = \int\! \diff x\,  (x \partial_x - p \partial_p) \Wc
    =  -p \partial_p \qmc - \int\! \diff x\,  \Wc \partial_x (x)
    = -  p \partial_p  \qmc - \qmc
\end{align}
where the Gaussian tails of $\Wc$ ensure no boundary term at infinity. Similar manipulations for $\qmh$ mean that
\begin{align}
  \partial_t \qmc &= -(1+p\partial_p)\qmc, \label{eq:qmc-dt}\\
  \partial_t \qmh &= \Bigl[\LLC^{(p)}_{D}-(1+p\partial_p)\Bigr]\qmh. \label{eq:qmh-dt}
\end{align}
during this third time step.
Applying Duhamel to the momentum dynamics (difference of \eqref{eq:qmh-dt} and \eqref{eq:qmc-dt} is $\LLC^{(p)}_D$) yields
\begin{align}\label{eq:term2z}
  \|\qmh_3-\qmc_3\|_{L^1}
  &\le (t_3-t_2)\,\sup_{t\in[t_2,t_3]}\bigl\|\LLC^{(p)}_{D}[\qmc_t]\bigr\|_{L^1}\\
   &= \tau_3 \,\frac{D}{2}\,\sup_{t\in[t_2,t_3]} \|\partial_p^2 \qmc_t\|_{L^1}.
\end{align}
Equation \eqref{eq:qmc-dt} is solved by exact scaling,
\[
  \qmc_t(p)=e^{-(t-t_2)}\,\qmc_2(e^{-(t-t_2)}p),
\]
so $\|\partial_p^2 \qmc_t\|_{L^1}=e^{-2(t-t_2)}\|\partial_p^2 \qmc_2\|_{L^1}$. Therefore the supremum over $t\in[t_2,t_3]$ in \eqref{eq:term2z} is attained at $t=t_2$, as expected:
\begin{align}\label{eq:term3}
  \|\qmh_3-\qmc_3\|_{L^1}
  &\le  \,\frac{\tau_3 D}{2}\,\|\partial_p^2 \qmc_2\|_{L^1}\\
  & = C_2\,\frac{\tau_3 D}{e^{-2\tau_3}}.
\end{align}
Here, using $\|\partial_p^2 \qmc_3\|_{L^1}=e^{-2\tau_3}\|\partial_p^2 \qmc_2\|_{L^1}$, we have defined 
\begin{align}
  C_2:=\frac{1}{2}e^{-2\tau_3}\,\|\partial_p^2 \qmc_2\|_{L^1} = \frac{1}{2} \|\partial_p^2 \qmc_3\|_{L^1} \approxx 0.3726
\end{align}
which, per \eqref{eq:q3}, is independent of $h$ for our choice of parameters.

\subsubsection{Final quantum bound}

Inserting \eqref{eq:term3} and \eqref{eq:term4} into the triangle inequality \eqref{eq:triangle} we get
\begin{align}\label{eq:quantum-final-bound}
	\|\qmo_3-\qmc_3\|_{L^1} 
	&\le  \|\qmo_3-\qmh_3\|_{L^1} + \|\qmh_3-\qmc_3\|_{L^1}\\
	&\le \left[C_1\frac{\tau_1 + \tau_2}{h e^{-2\tau_1}} + C_2\,\frac{\tau_3 }{e^{-2\tau_3}}\right]D\\
	&\eqq \frac{C_1\left(1+\tfrac{1}{6}\log h^{-1}\right) + C_2\tfrac{2}{3}\log h^{-1} }{h^{\frac43}}D\\
    &\le
    C_{\mathrm{qu}}\frac{1+ \log h^{-1}}{ h^{\frac43} }D
\end{align}
for positive constant
\begin{align}
    C_{\mathrm{qu}}:= C_1+ \frac{2}{3}C_2 \approxx 3.846
\end{align}
Hence
\begin{align}\label{eq:q-thresh}
\|\qmo_3-\qmc_3\|_{L^1} \to 0 \quad\text{if}\quad D \eqq o\!\left(\frac{h^{\frac43}}{1+\log h^{-1}}\right).
\end{align}

\subsection{Classical diffusion error}

The classical case is much the same. We seek to bound $\|\cmo_3-\cmc_3\|_{L^1}$, where $\cmo(p):=\int\! \fo(x,p)\, d x$ is the classical momentum distribution in the presence of diffusion.  We start by considering the Duhamel bound on the overall state error $\|\fo_t-\fc_t\|_{L^1}$ for arbitrary time $t$.

Let $\fc_t$ be the solution to the classical closed-system dynamics
\begin{align}
	\frac{\partial\fc_t}{\partial t} = {\LLC}_H[\fc_t] = - (\partial_x \fc_t )(\partial_p H(t) )+ (\partial_p \fc_t)( \partial_x H(t)) 
\end{align}
as described in the first section, where $H=H(t)$ is the classical three-bump Hamiltonian given by \eqref{eq:ham}.  We will compare it to the classical open-system trajectory $\fo_t$, starting from the same initial pure Gaussian state \eqref{eq:initial-state}, which solves
\begin{align}
	\frac{\partial\fo_t}{\partial t} = {\LLC}[\fo_t] = {\LLC}_H[\fo_t]+{\LLC}_D[\fo_t].
\end{align}
The contribution from diffusion 
\begin{align}
	{\LLC}_D[\fo] &= \frac{1}{2} D (\partial_x^2\fo + \partial_p^2\fo ),
\end{align}
is the identical differential operator as the quantum case, \eqref{eq:diff-superop}.
We want to show that introducing diffusion does not much change the final momentum distribution $\cmo_3(p) = \int\! \diff x\, \fo_3(x,p)$, i.e., that $\|\cmo_3-\cmc_3\|_{L^1}$ is small.

\subsubsection{Duhamel bound on growth of overall state error \texorpdfstring{$\|\fo_t-\fc_t\|_{L^1}$}{}}

By Duhamel,
\begin{align}
	\fo_t-\fc_t = \int_0^t \! ds \, e^{(t-s) {\LLC}} \left({\LLC}-\LLC_H\right)[\fc_s]
\end{align}
we have
\begin{align}
	\|\fo_t-\fc_t\|_{L^1} 
	&\le \int_0^t \! ds \, \| {\LLC}_D[\fc_s] \|_{L^1} \\
	&= \frac{D}{2} \int_0^t \! ds \, \left(\|\partial_x^2 \fc_s\|_{L^1} + \|\partial_p^2 \fc_s\|_{L^1}\right),
\end{align}
since $e^{(t-s)\LLC}$ is an $L^1$-contraction.  As before, the error in the overall (now classical) state will become large during the third step. So we introduce an intermediate classical trajectory $\fh_t$ that is equal to the closed-system evolution $\fc_t$ up to the end of the second step  ($\fh_2=\fc_2$) and which then experiences the diffusion.  We will bound $\|\cmo_3 - \cmc_3\|_{L^1}$ by starting with the triangle inequality with the intermediate trajectory
\begin{align}\label{eq:trianglec}
	\|\cmo_3-\cmc_3\|_{L^1}  &\le  \|\cmo_3-\cmh_3\|_{L^1} + \|\cmh_3-\cmc_3\|_{L^1} .
\end{align}

\subsubsection{Bounding overall state error \texorpdfstring{$\|\cmo_3-\cmh_3\|_{L^1} \le  \|\fo_2-\fc_2\|_{L^1}$}{} through end of step 2}

For most of this subsection we will not denote time-dependence explicitly. Throughout all evolution, the closed-system classical density has the conditional-Gaussian form
\begin{align}
\fc(x,p)=\gauss_{\sigma_x}(x)\,\gauss_{\sigma_p}(p-r x^2),\qquad\qquad \mathrm{with}\qquad \gauss_\sigma(u):=(2\pi\sigma^2)^{-\frac12}\exp[-u^2/(2\sigma^2)],
\end{align}
where $\sigma_x,\sigma_p>0 $ and $r\ge 0$ have implicit time dependence. 
Define the dimensionless variables
\begin{align}
y:=x/\sigma_x,\quad z:=(p-rx^2)/\sigma_p
\end{align}
so that $y,z$ are i.i.d.\ standard Gaussians.  Thus for any function $Q$
\begin{align}
\int \diff x \int \diff p \, Q(y,z) \fc(x,p) = \int dy \int dz \, Q(y,z) \gauss_1(y)\gauss_1(z) 
= \mathbb{E} \, Q(y,z)
\end{align}
where $\mathbb{E}$ denotes the expectation with respect to $y,z\stackrel{\text{iid}}{\sim} \mathcal{N}(0,1)$.

It will be cleaner to work with the derivatives of logs of Gaussians, so we note that for any smooth function $h>0$,
\begin{align}\label{eq:log-deriv}
\partial^2 h=h\left[\left(\partial\log h\right)^2+\partial^2\log h\right].
\end{align}
First we compute the $p$-derivatives. Since 
\begin{align}
\partial_p\log \gauss_{\sigma_p}(p-rx^2)=-(p-rx^2)/\sigma_p^2=-z/\sigma_p,
\end{align}
we have
\begin{align}
\partial_p^2 \fc=\sigma_p^{-2}(z^2-1)\,\fc.
\end{align}
Taking the $L^1$ norm gives the exact identity
\begin{align}\label{eq:dp2f-bound}
\|\partial_p^2 \fc\|_{L^1}
&= \int \diff x \int \diff p \, \left|\partial_p^2 \fc(x,p)\right|\\
&= \sigma_p^{-2}\int \diff x \int \diff p \, |z^2-1|\fc(x,p)\\
&=\sigma_p^{-2}\,\mathbb{E}\!\left|z^2-1\right|\\
&=C_3\,\sigma_p^{-2}, 
\end{align}
with constant
\begin{align}
	C_3 := \mathbb{E} |z^2-1| 
	= 2\sqrt{\frac{2}{\pi e}}
	\approx 0.9679
\end{align}

Now we compute the $x$-derivatives
\[
\partial_x\log \fc
=\partial_x\log \gauss_{\sigma_x}(x)+\partial_x\log \gauss_{\sigma_p}(p-rx^2)
=-\frac{x}{\sigma_x^2}+\frac{2r x}{\sigma_p^2}(p-rx^2) 
=\frac{y}{\sigma_x}\left(-1+2 b z\right). 
\]
with dimensionless (non-random) parameter
\[
b:=r\,\sigma_x^2\,\sigma_p^{-1}\ge 0
\]
Differentiating once more,
\[
\partial_x^2\log \fc=-\frac{1}{\sigma_x^2}+\frac{2r}{\sigma_p^2}\bigl(p-3rx^2\bigr)
=-\frac{1}{\sigma_x^2}+\frac{2b}{\sigma_x^2}z-\frac{4b^2}{\sigma_x^2}y^2. 
\]
Combining these using \eqref{eq:log-deriv} we get 
\begin{align}
\partial_x^2 \fc
=\frac{\fc}{\sigma_x^2}\Bigl[
y^2-1+2b z\,(1-2y^2)+4b^2 y^2\,(z^2-1)
\Bigr].
\end{align}
so
\begin{align}
\|\partial_x^2 \fc\|_{L^1}
=\frac{1}{\sigma_x^2}\,
\mathbb{E}\Bigl|\,y^2-1+2b z(1-2y^2)+4b^2 y^2(z^2-1)\Bigr|.
\end{align}
Apply the triangle inequality and independence of $y,z$:
\begin{align}
\|\partial_x^2 \fc\|_{L^1}
\le \frac{1}{\sigma_x^2}\Big[
\mathbb{E}\left|y^2-1\right|
+2b\,\mathbb{E}\left|z(1-2y^2)\right|
+4b^2\,\mathbb{E}\left[y^2\right]\, \mathbb{E}\left|z^2-1\right|
\Big]. 
\end{align}
Since $\mathbb{E}[y^2]=1$, this becomes
\begin{align}\label{eq:dx2f-bound}
\|\partial_x^2 \fc\|_{L^1}
\le \frac{1}{\sigma_x^2}\left[(1+4b^2)C_3+2b\,C_4\right]
\end{align}
with constant
\begin{align}
	C_4 := \mathbb{E} |z (1 - 2 y^2)|
	= \sqrt{\frac{2}{\pi }} \left[1+\frac{4 e^{-\frac14}}{\sqrt{\pi }}-2 \, \mathrm{erf}\left(\frac{1}{2}\right)\right]
	\approx 1.370.
\end{align}
In the original parameters ($b=r\sigma_x^2\sigma_p^{-1}$),
\begin{align}
\|\partial_p^2 \fc\|_{L^1}=C_3\,\sigma_p^{-2},\qquad
\|\partial_x^2 \fc\|_{L^1}
\le C_3\bigl(\sigma_x^{-2}+4r^2\sigma_x^2\sigma_p^{-2}\bigr)
+2C_4\,r\,\sigma_p^{-1}.
\end{align}
Therefore 
\begin{align}
	\|\cmo_3-\cmh_3\|_{L^1} &\le \|\fo_2-\fh_2\|_{L^1} 
    \\
	&\le \|\fo_2-\fc_2\|_{L^1} \\
	&= \frac{D}{2} \int_0^{t_2} \! ds \, \left(\|\partial_x^2 \fc_s\|_{L^1} + \|\partial_p^2 \fc_s\|_{L^1}\right),\\
	&\le \frac{t_2 D}{2}\sup_{t\in [0,t_2]} \left(\|\partial_x^2 \fc_t\|_{L^1} + \|\partial_p^2 \fc_t\|_{L^1}\right)\\
    &\le \frac{(\tau_1+\tau_2) D}{2}\sup_{t\in [0,t_2]} \left[C_3 (\sigma_x^{-2}+\sigma_p^{-2}+4 r^2 \sigma_x^{2} \sigma_p^{-2})  + 2 C_4 r \sigma_p^{-1}\right]\label{eq:c3c4insert}\\
	&\le \frac{(\tau_1+\tau_2) D}{2}\left[\sup_{t\in [0,t_2]} \left[C_3 (\sigma_x^{-2}+\sigma_p^{-2})\right]+\sup_{t\in [0,t_2]} \left[4 C_3 r^2 \sigma_x^{2} \sigma_p^{-2}+2 C_4 r \sigma_p^{-1}\right]\right]\\
    &\le \frac{(\tau_1+\tau_2) D}{2}\left[C_3 (h^{-1}e^{-2\tau_1}+h^{-1}e^{2\tau_1})+4 C_3  \frac{\tau_2^2} {1+2\tau_2^2} e^{4\tau_1}  + 2 C_4 \frac{\tau_2}{\sqrt{ 1+2\tau_2^2}} h^{-\frac12}e^{\tau_1}\right]\label{eq:t2sup2}\\
    &\le \frac{(\tau_1+\tau_2) D}{h e^{-2\tau_1}}\left[C_3 \frac{e^{-4\tau_1}+1}{2}+2 C_3  \frac{\tau_2^2} {1+2\tau_2^2} h e^{2\tau_1}  + C_4 \frac{\tau_2}{\sqrt{ 1+2\tau_2^2}} h^{\frac12}e^{-\tau_1}\right]\\
     &\le \frac{(\tau_1+\tau_2) D}{h e^{-2\tau_1}}\left[
     C_3  \frac{1+4\tau_2^2} {1+2\tau_2^2}  
     + C_4 \frac{\tau_2}{\sqrt{ 1+2\tau_2^2}} \right]\label{eq:term8}
\end{align}
In \eqref{eq:c3c4insert} we have used $b=r(t)\sigma_x^2\sigma_p^{-1}$ and the bounds \eqref{eq:dp2f-bound} and \eqref{eq:dx2f-bound} on $\partial_p^2 \fc$ and $\partial_x^2 \fc$. (Recall that $b, r, \sigma_x, \sigma_p$ all have time dependence.) 
In \eqref{eq:t2sup2} we have used $r(t)=0$ for $t\in[0,t_1]$ and, for $t\in[t_1,t_2]$,
\[
r(t)=\int_{t_1}^{t}\chi_2(s)\,ds,
\qquad\text{so that}\qquad
0\le r(t)\le \int_{t_1}^{t_2}\chi_2(s)\,ds=\tau_2.
\]
Thus the relevant supremands are maximized at the endpoints $r=0$ (attained at $t=t_1$) or $r=\tau_2$ (attained at $t=t_2$).
In \eqref{eq:term8} we use $h\le 1$, and our technical assumption that $0< \tau_1 < \tfrac{1}{4}\log h^{-1}$.

\subsubsection{Bounding momentum error \texorpdfstring{$\|\cmh_3-\cmc_3\|_{L^1}$}{} during step 3}

As for the Wigner function in the quantum case, the dynamics for the classical phase-space distribution during the third step separate into terms involving only $x$ and only $p$:
\begin{align}
	\partial_t \fc &= \LLC^{(x)}_{3,H} [\fc] + \LLC^{(p)}_{3,H} [\fc],\\
	\partial_t \fh &= \left(\LLC^{(x)}_{3,H} + \LLC^{(x)}_D\right) [\fh] + \left(\LLC^{(p)}_{3,H} + \LLC^{(p)}_D\right) [\fh],
\end{align}
Thus, for the same reasoning as Eqs.~\eqref{eq:qmc-dt} and \eqref{eq:qmh-dt}, we can write the dynamics for the classical momentum distributions $\cmh(p) := \int\! \diff x\, \fh(x,p)$ and $\cmc(p) := \int\! \diff x\, \fc(x,p)$ as
\begin{align}
    \partial_t \cmc &= -(1 +p \partial_p ) \cmc, \label{eq:cmc-dynamics} \\
    \partial_t \cmh &= 
    \left[\LLC^{(p)}_{D}-  (1 +p \partial_p)\right] \cmh 
\end{align}
during the third time step. We use Duhamel and the fact that the stretching $p\mapsto e^{\tau_3} p$ weakens $\partial_p^2$ monotonically:
\begin{align}
	\|\cmh_3-\cmc_3\|_{L^1}
	&\le (t_3-t_2) \sup_{t\in [t_2,t_3]}  \left\| {\LLC}^{(p)}_D[\cmc_t] \right\|_{L^1} \\
	&=  \frac{\tau_3 D}{2}\, \sup_{t\in [t_2,t_3]}  \left\| \partial_p^2 \cmc_t \right\|_{L^1} \\
	&= \frac{\tau_3 D}{2}\,\|\partial_p^2 \cmc_2\|_{L^1} \\
	\label{eq:term2c}
	&= C_5\, \frac{\tau_3 D}{2e^{-2\tau_3}},
\end{align}
where, similar to the quantum case, we use the fact that \eqref{eq:cmc-dynamics} is solved by just scaling the distribution at $t=t_2$, so that $\|\partial_p^2 \cmc_3 \|_{L^1}=e^{-2\tau_3}\|\partial_p^2 \cmc_2\|_{L^1}$.  Here we have defined
\begin{align}
	C_5:=\|\partial_p^2 \cmc_3\|_{L^1} \approxx 0.60002,
\end{align}
which is independent of $h$ for our choice of parameters, as seen from the expression \eqref{eq:cmc3} for the final classical momentum distribution $\cmc_3$.

\subsubsection{Final classical bound}

Inserting \eqref{eq:term8} and \eqref{eq:term2c} into the triangle inequality \eqref{eq:trianglec}
\begin{align}
	\|\cmo_3-\cmc_3\|_{L^1}
	&\le  \|\cmo_3-\cmh_3\|_{L^1} + \|\cmh_3-\cmc_3\|_{L^1} \\
	&\le \left[
	C_3  \frac{1+4\tau_2^2} {1+2\tau_2^2}  
	+ C_4 \frac{\tau_2}{\sqrt{ 1+2\tau_2^2}} \right] \frac{(\tau_1+\tau_2) D}{ h e^{-2\tau_1}} + \frac12 C_5\, \frac{\tau_3 D}{e^{-2\tau_3}}\\
    &\lee  C_{\mathrm{cl}}\frac{1+\log h^{-1}}{h^{\frac43}} D\label{eq:classical-final-bound}
\end{align}
for positive constant
\begin{align}
	C_{\mathrm{cl}} &:=
	C_3  \frac{1+4\tau_2^2} {1+2\tau_2^2}  
	+ C_4 \frac{\tau_2}{\sqrt{ 1+2\tau_2^2}} + \frac12 C_5 \approxx 2.704.
\end{align}
Hence
\begin{align}
\|\cmo_3-\cmc_3\|_{L^1} \to 0 \quad\text{if}\quad D=o\!\left(\frac{h^{\frac43}}{1+\log h^{-1}}\right),
\end{align}
matching the asymptotic quantum threshold \eqref{eq:q-thresh}.

\section{Final overall bound}\label{sec:overall-bound}

Combining \eqref{eq:classical-final-bound} and \eqref{eq:quantum-final-bound} with yet another triangle inequality, we have 
\begin{align}\label{eq:final-bound}
    \|\cmo_3-\qmo_3\|_{L^1} &\ge \|\cmc_3-\qmc_3\|_{L^1} - \|\cmo_3-\cmc_3\|_{L^1} - \|\qmo_3-\qmc_3\|_{L^1} \\
    &\ge \bar{c} - C \frac{1+\log h^{-1}}{h^{\frac43}} D
\end{align}
with positive constants
\begin{alignat}{2}\label{eq:final-constants}
    \bar{c}&:=\|\cmc_3-\qmc_3\|_{L^1} & &\approxx 0.2726,\\
    C&:=   C_{\mathrm{cl}} +  C_{\mathrm{qu}}  & & \approxx 6.550.
\end{alignat}
We can then conclude that the open-system discrepancy is bounded as
\begin{align}
|\langle \hat{\gp}_0\rangle_{\rhoo_3} - \langle \gp_0\rangle_{\fo_3}| &= |\langle \gp_0\rangle_{\qmo_3} - \langle \gp_0\rangle_{\cmo_3}|\\
&\ge c_0-C \frac{1+\log h^{-1}}{h^{\frac43}} D
\end{align}
where $\gp_0 := \exp(-p^2)$, $\hat{\gp}_0 := \exp(-\hat{p}^2)$, and the closed-system discrepancy is 
\begin{align}
c_0:=|\langle \hat{\gp}_0\rangle_{\rhoc_3} - \langle \gp_0\rangle_{\fc_3}| = |\langle \gp_0\rangle_{\qmc_3} - \langle \gp_0\rangle_{\cmc_3}| \approxx 0.06412.
\end{align}

\bibliographystyle{unsrt}
\bibliography{references}

\end{document}